%% file: MooreReadPaper.tex
\documentclass[aps,prX,reprint,showpacs,superscriptaddress]{revtex4-1}
\pdfoutput=1
\usepackage{graphicx}
\usepackage{amsmath}
\usepackage{amsfonts}
\usepackage{color}
\usepackage{hyperref}
\usepackage{multirow}
\usepackage[usenames,dvipsnames]{xcolor}
\usepackage{subfigure}

\newcommand{\Ket}[1]{| #1 \rangle}

\newcommand{\Braket}[1]{\langle #1 \rangle}

\newcommand{\MPQ}{Max-Planck-Institut f\"ur Quantenoptik, Hans-Kopfermann-Str.\ 1, D-85748 Garching, Germany}
\newcommand{\madrid}{Instituto de F\'isica Te\'orica, UAM-CSIC, Madrid, Spain}
\newcommand{\princeton}{Department of Physics, Princeton University, Princeton, NJ 08544, USA}

\makeatletter
\newcommand{\redboxed}[1]{\textcolor{red}{%
  \fbox{\normalcolor\m@th$\displaystyle#1$}}}
\makeatother

\begin{document}

% \date{March XY, 2015}

\title{Exact parent Hamiltonians of bosonic and fermionic Moore-Read states\\ on lattices and local models}

\begin{abstract}
We introduce a family of strongly-correlated spin wave functions on arbitrary spin-$1/2$ and spin-$1$ lattices in one and two dimensions. These states are lattice analogues of Moore-Read states of particles at filling fraction $1/q$, which are non-Abelian Fractional Quantum Hall states in 2D. One parameter enables us to perform an interpolation between the continuum limit, where the states become continuum Moore-Read states of bosons (odd q) and fermions (even q), and the lattice limit. We show numerical evidence that the topological entanglement entropy stays the same along the interpolation for some of the states we introduce in 2D, which suggests that the topological properties of the lattice states are the same as in the continuum, while the 1D states are critical states. We then derive exact parent Hamiltonians for these states on lattices of arbitrary size. By deforming these parent Hamiltonians, we construct local Hamiltonians that stabilize some of the states we introduce in 1D and in 2D.
\end{abstract}

\pacs{11.25.Hf, 73.43.-f, 75.10.Jm}

\author{Ivan Glasser}
\affiliation{\MPQ}
\author{J. Ignacio Cirac}
\affiliation{\MPQ}
\author{Germ\'an Sierra}
\affiliation{\madrid}\affiliation{\princeton}
\author{Anne E. B. Nielsen}
\affiliation{\MPQ}

\maketitle

\section{Introduction}
\label{SEC:Introduction}
The Fractional Quantum Hall (FQH) effect is one of the most fascinating phenomena in strongly correlated electronic systems, in which the electrons of a two-dimensional electron gas subject to a strong magnetic field form an incompressible quantum liquid supporting fractionally charged quasiparticle excitations. The understanding of this paradigm of topological order was in large part made possible by the discovery of analytical wave functions, such as the Laughlin's wave function \cite{Laughlin1983}, describing the electrons in a partially filled Landau level. 

Since its discovery in 1987 \cite{Willett1987}, one FQH state has attracted a lot of attention : Unlike the states at filling factors with odd denominators, the $\nu=5/2$ FQH state with electrons occupying the second Landau level cannot be explained by a hierarchical construction based on the Laughlin's states \cite{Haldane1983}. This opens the door to the possibility of electron pairing and emergence of non-Abelian quasiparticle excitations. Indeed the leading candidate for the description of the $\nu=5/2$ FQH state is the Moore-Read ``Pfaffian'' state at filling $1/2$ \cite{Moore1991,Greiter1991,Greiter1992}, describing the wave function of the electrons in the second Landau level. Moore-Read states have a wave function defined by \cite{Moore1991}
\begin{align}
\label{MooreRead}
\psi(w_1, \ldots, w_M)\propto\prod_{i<j}(w_i-w_j)^{q}\text{Pf}\left[\frac{1}{w_i-w_j}\right]e^{-\frac{1}{4} \sum_i |w_i|^2},
\end{align}
where $w_i$ are the positions of the particles on the complex plane, $1/q$ is the filling factor and the magnetic length has been set to one. They support fractionally charged non-Abelian anyons possessing Majorana fermion states at zero energy  \cite{Read2000,Ivanov2001}. These non-Abelian anyons have attracted a lot of attention due to their applications to topological quantum computation \cite{Kitaev20032,Nayak2008}.

There has been a lot of recent interest in finding FQH physics in other systems. Indeed, it is a fundamental problem to understand the origin and properties of states supporting non-Abelian anyons. Finding different systems exhibiting the same physics is an important step towards an explanation of these phenomena, for which our understanding is still far from complete. Moreover, experimental realizations of FQH states and manipulation of their quasiparticles are a challenge. As such it is interesting to search for new possibilities for realising FQH physics and variations thereof experimentally. In particular, lattice models without Landau levels may pave the way towards experimental realization and investigation of FQH-like states, for example in optical lattices.

Two main approaches have been followed to recreate the FQH effect in lattices. In the first approach, one tries to mimic electrons in a magnetic field by replacing the fractionally filled Landau level by a nearly flat fractionally filled Chern band and adding local interactions \cite{Sheng2011,Neupert2011,Wang2011,Sun2011,Regnault2011}. Non-Abelian FQH states where found in such lattice models with topological flat bands \cite{Wang2012,Liu2013,Liu2013a,BERGHOLTZ2013}.

In the second approach, instead of trying to reproduce the interactions between the electrons and a magnetic field, the focus is on reproducing the FQH wave functions on lattices. This approach first started with the introduction by Kalmeyer and Laughlin of the bosonic Laughlin state at filling fraction $1/2$ on a square lattice \cite{Kalmeyer1987}, for which a parent Hamiltonian was later derived on the torus \cite{Schroeter2007,Thomale2009} and for more general lattices in the thermodynamic limit \cite{Nielsen2012,Greiter2014,Tu2014} (see also \cite{Nielsen2013,Bauer2014} for related results). Hamiltonians were also derived for non-Abelian chiral spin liquids with excitations with $SU(2)_k$ statistics \cite{Greiter2014}, but only in the thermodynamic limit. For $k=2$ this corresponds to a bosonic lattice Moore-Read state at filling fraction 1 on a spin-$1$ lattice \cite{Greiter2009}. Moore-Read states of bosons have also been considered on one dimensional lattices, were parent Hamiltonians have been obtained \cite{Paredes2012}. Other filling fractions and Moore-Read states on spin-$1/2$ lattices have not been introduced before.

In the continuum, a useful description of FQH states \cite{Moore1991} uses wave functions expressed in terms of correlators of the related edge Conformal Field Theory (CFT). This description was extended to lattice systems in one and two dimensions \cite{cftimps,Nielsen2012} by writing the wave functions of spin systems as CFT correlators. The states obtained can be seen as an infinite dimensional version of Matrix Product States (MPS) which gives a unified treatment of 1D and 2D lattices systems. An alternative way of implementing MPS ideas to 2D FQH systems has been developed in \cite{Estienne2013}. For the Laughlin states, this construction of wave functions written as CFT correlators provides states that are close to, but not exactly the same as the Kalmeyer-Laughlin states on a lattice of finite size, but that become the same in the thermodynamic limit \cite{Nielsen2012}. This modification of the lattice construction has made it possible to construct exact parent Hamiltonians for strongly interacting lattice spin systems of arbitrary sizes \cite{Nielsen2011,Tu2013,Bondesan2014483,Tu2014328,Nielsen2012}, including topological FQH states such as Laughlin states of hardcore bosons and fermions \cite{Tu2014}. It was shown in some cases that these states could be stabilized by a local Hamiltonian \cite{cftimps,Tu2014328,Nielsen2013,Glasser2014}. This description has been applied to find a Hamiltonian for the $SU(2)_2$ spin models but the focus was on the one dimensional spin chain \cite{Nielsen2011} and the corresponding two dimensional non-Abelian FQH state has not been analyzed so far.

In this paper we fill this gap by extending the construction of lattice wave functions from CFT correlators to non-Abelian FQH states. Using this approach we construct a family of lattice versions of the Moore-Read state at filling fraction $1/q$. This family of states allows us to interpolate between the continuum limit, where all states become continuum Moore-Read wave functions, and a lattice limit. The states are defined on arbitrary spin-$1/2$ and spin-$1$ lattices in one or two dimensions, where the value of each spin can be mapped to an occupation number of bosonic (odd q) or fermionic (even q) particles. We provide numerical evidence that the states are critical states in one dimension and that in two dimensions they are topologically ordered states. It is shown that states defined on a spin-$1/2$ square lattice (for $q=2$) and states defined on a spin-$1$ square lattice (for $q=1$) have a topological entanglement entropy which is constant along the interpolation to the continuum. This suggests that the lattice states have the same topological properties as Moore-Read states in the continuum. 

Since we use an approach based on the wave functions, it is then relevant to ask whether these states are ground states of physical Hamiltonians. Using properties from CFT, we derive parent Hamiltonians for the wave functions in the lattice limit. These parent Hamiltonians have long-range interactions, are exact on any lattice of arbitrary size and we find numerically that they have a non-degenerate ground space for $q\leq2$. By deforming these parent Hamiltonians, we then show numerical evidence that the state at filling fraction 1 on a spin-1 square lattice can be stabilized by a local Hamiltonian in one and two dimensions, while the state at filling fraction $1/2$ on a spin-$1/2$ square lattice can be stabilized by a local Hamiltonian in one dimension, which is a first step towards an experimental realization of these states.

The paper is organized as follows : In Sec.~\ref{section1}, lattice Moore-Read states are defined from correlators of conformal fields. It is shown in Sec.~\ref{section2} that these states reduce to Moore-Read states of particles in the continuum. Properties of the states are computed in Sec.~\ref{section3}, where evidence that the states are critical in one dimension and that the topological properties remain the same along the interpolation between the continuum and the lattice states in two dimensions is presented. Parent Hamiltonians are derived in Sec.~\ref{section4}. Finally Sec.~\ref{section5} provides numerical evidence that some of these Hamiltonians can be deformed into local Hamiltonians stabilizing the lattice states.

\section{Lattice States From Correlators Of Conformal Fields}
\label{section1}

Let us consider a lattice with $N$ sites at positions $z_j$, $j\in\{1, 2, \ldots , N\}$ in the complex plane. We will refer to one dimensional models when the $z_j$ are restricted to the unit circle and two dimensional models otherwise. Let $a$ be the average area per site (in 1D $a$ is the average distance between two sites on the unit circle) and $\eta\equiv\frac{a}{2\pi}$ a positive real number. The local basis at site $j$ is $\Ket{n_j}$, where $n_j$ is an integer that will be interpreted as the number of particles at site $j$. We consider two classes of models, labelled by $S\in{\{\frac{1}{2},1\}}$ :
\begin{itemize}
\item Models with $S=\frac{1}{2}$ are defined on a Hilbert space of size $2^N$ : the two states in the local basis are $n_j\in\{0,1\}$ and correspond to the absence/presence of a particle at site $j$. These states can be expressed in terms of spin $\frac{1}{2}$ variables at each site : $s_i=2 n_i-1$.
\item Models with $S=1$ are defined on a Hilbert space of size $3^N$ : the local basis is $n_j\in \{0,1,2\}$, which we interpret as the presence of $0/1/2$ particles at site $j$. These states can be expressed in terms of spin $1$ variables at each site : $s_i=n_i-1$.
\end{itemize}

In general, a wave function defined on one of these two spaces will have the form
\begin{align}
\Ket{\psi}_S &= \sum_{n_1, \dots, n_N} \psi(n_1, \ldots, n_N) \Ket{n_1, \dots, n_N}_S.
\end{align}
In the following, we will consider wave functions for which the coefficients $\psi(n_1, \ldots, n_N)$ can be expressed as the correlator of some conformal operators.
Let us introduce the operators
\begin{align}
V_{n_{j}}(z_{j})&=\chi(z_{j})^{\bar{\delta}_{n_j}} :e^{i(qn_j-\eta)\phi (z_{j})/\sqrt{q}}:,
\end{align}
where $\phi(z)$ is a chiral bosonic field from the $c=1$ Conformal Field Theory (CFT), $\chi$ is a Majorana fermion field, $:\ldots:$ denotes normal ordering, $q$ is a positive integer and $\bar{\delta}_{n_j}$ is 1 if $n_j=1$ and 0 otherwise. We also define a phase coefficient
\begin{align}
U_{n_{j}}(z_{j})&=\xi_j^{n_j} e^{i\pi (j-1) \eta n_j} ,
\end{align}
where the $\xi_j$ are phase factors to be specified.

We propose to consider the wave functions defined by the correlator of the previous operators :
\begin{align}
\psi_S(n_1, \ldots, n_N)&\propto \Braket{\mathcal{V}_{n_1}(z_1) \dots \mathcal{V}_{n_N}(z_N)},
\end{align}
where 
\begin{align}
\mathcal{V}_{n_j}(z_j)=U_{n_{j}}(z_{j})V_{n_{j}}(z_{j}). \label{defvertex}
\end{align}
These wave functions are labelled by three parameters : $q$, $\eta$, $S$ and will be referred to as the $(q,\eta)_S$  CFT states.

\begin{figure}[htb]
\centering
\begin{scriptsize}
\def\svgwidth{8.6cm}
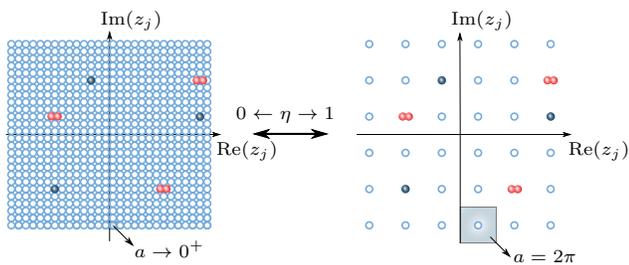
\caption{\label{fig:Interpolation}
  Illustration of a square lattice on the complex plane in the continuum limit ($\eta\rightarrow 0, N\rightarrow \infty$) and in the lattice limit($\eta\rightarrow 1$). At each site there can be $0$ (blue circle), $1$ (blue disk) or $2$ (red disks) particles. The interpolation is performed by fixing the number of particles $M=\eta \frac{N}{q}$ and by varying $\eta=\frac{a}{2\pi}$ between $0$ and $1$, which changes the number of lattice sites per particle between infinity and $q$. 
}
\end{scriptsize}
\end{figure}

Let us now evaluate the correlator. Note that we do not need to add a background charge in this setting. The correlator is zero unless $\sum_{i=1}^N n_i=\eta N/q$. This condition fixes the total number of particles in the system to $M\equiv\sum_{i=1}^N n_i=\eta N/q$. $\eta/q$ is therefore the lattice filling fraction and $\eta$ is a parameter which interpolates between the continuum limit ($\eta \rightarrow 0$, $N\rightarrow \infty$, number of particles conserved) with infinitely many lattice sites per particle and the lattice limit ($\eta=1$) at which the lattice filling fraction is equal to $1/q$, which corresponds to the Landau level filling fraction in the FQH effect (Fig. \ref{fig:Interpolation}). Let us write $\delta_n=1$ if $\sum_{i=1}^N n_i=M$ and $\delta_n=0$ otherwise. The evaluation of the correlator yields \cite{cft}

\begin{align}
\label{CFTstate}
\psi_S(n_1, \ldots, n_N)\propto &\delta_n \prod_{i<j}(z_i-z_j)^{q n_i n_j} \nonumber\\
&\times \text{Pf}_{n_i=n_j=1}\left[\frac{1}{z_i-z_j}\right]\prod_l f_N(z_l)^{n_l},
\end{align}
where $f_N(z_{l})\equiv \xi_{l}\prod_{j(\neq l)}(z_{l}-z_{j})^{-\eta}$ and the Pfaffian is evaluated at the coordinates where $n_i=1$. The Pfaffian is antisymmetric, so these are bosonic states when $q$ is odd and fermionic states when $q$ is even. Note that this formula defines different states for $S=1/2$ than for $S=1$, the difference being that in the second case the $n_i$ can take the value $2$. The states $(q,\eta)_{1/2}$ are therefore projections of the states $(q,\eta)_{1}$ onto the Hilbert space allowing only for single occupancy at each site, while states with $S=1$ and $q$ odd (resp. even) can have sites with two bosons (resp. fermions of different types). The state $(q=1,\eta=1)_{1/2}$ is trivial since the number of particles is fixed to $N$ and there are $N$ sites, so in the following we restrict ourselves to states $(q,\eta)_{1/2}$ with $q\geq2$ and $(q,\eta)_{1}$ with $q\geq1$ (Table \ref{table:CFTstates}).

\begin{table}[htb]
  \caption{\label{table:CFTstates}
  First $(q,\eta)_S$ CFT states }
\centering
\begin{tabular}{rr|c|c}
\hline
\hline
%\hline
\multicolumn{4}{c}{Lattice limit $\eta=1$}\\
\hline
& \multirow{2}{*}{$S$} & \multirow{2}{*}{$\frac{1}{2}$}  &  \multirow{2}{*}{$1$} \\
q& &  & \\
\hline
$1$ &  & $\times$ & bosonic $SU(2)_2$ $(q=1,\eta=1)_{1}$  \\
$2$ &  & fermionic $(q=2,\eta=1)_{1/2}$ & fermionic $(q=2,\eta=1)_{1}$ \\
$\vdots$ &  & $\vdots$ & $\vdots$ \\
\hline
\hline
%\hline
\multicolumn{4}{c}{Continuum limit $\eta\rightarrow0^{+}$}\\
\hline
& \multirow{2}{*}{$S$} & \multirow{2}{*}{$\frac{1}{2}$}  &  \multirow{2}{*}{$1$} \\
q& &  & \\
\hline
$1$ &  & $\times$ & bosonic Moore-Read \\
$2$ &  & fermionic Moore-Read & fermionic Moore-Read \\
$\vdots$ &  & $\vdots$ & $\vdots$ \\
\hline
\hline
\end{tabular}

\end{table}

\section{The CFT states become Moore-Read states in the Continuum limit}
\label{section2}

In this section we consider a two dimensional lattice defined on a disk $\mathcal{D}$ of radius $R\rightarrow \infty$ and show that the CFT states we have introduced reduce to Moore-Read states of particles in the continuum, that is Eq.~(\ref{MooreRead}),  when $\eta\rightarrow 0, N\rightarrow \infty$ and the number of particles $M$ is fixed. We restrict ourselves to lattices where the area per site $a_i$ is constant equal to a, but the derivation remains true for any lattice if we make $\eta$ position dependent \cite{Tu2014}.

Let us first compute $\prod_l f_N(z_l)^{n_l}$. Notice that $|f_N|=\exp(-\sum_{j(\neq l)}\eta \ln(|z_{l}-z_{j}|))$ and since $\eta=\frac{a}{2\pi}$, in the continuum limit this sum can be replaced by an integral $\int_{\mathcal{D}} \ln(|z_{l}-z|)) dz^2/2\pi$. This integral evaluates, in the thermodynamic limit, to $|z_l|^2+\text{constant}$ \cite{Moore1991}, so that 
\begin{align}
f_{N\rightarrow \infty}(z_l)\propto \xi_l  e^{-i g_l} e^{-|z_l|^2/4},
\end{align}
where $g_l=\text{Im}\left(\sum_{j(\neq l)}\eta \ln(z_{l}-z_{j})\right)$ is a real number. It was found numerically in Ref.~\cite{Tu2014} and Ref.~\cite{Nielsen2012} that this formula was an accurate approximation even for moderately large N. We thus get that
\begin{align}
\prod_l f_{N\rightarrow \infty}(z_l)^{n_l} \propto \left( \prod_l \xi_l^{n_l}  e^{-i n_l g_l} \right) e^{-\frac{1}{4} \sum_l |z_l|^2 n_l}.
\end{align}
In the rest of this section we set the phase factors such that $\xi_l=e^{i g_l}$ to get rid of the overall gauge factor, which does however not change properties like the particle-particle correlation function and entanglement entropy of the state.

\subsection{Continuum limit of the $S=\frac{1}{2}$ states}

Let us now write the complete wave function in the continuum limit :
\begin{align}
\label{CFTcontinuum}
\psi_{S}(n_1, \ldots, n_N)\propto &\delta_n \prod_{i<j}(z_i-z_j)^{q n_i n_j} \nonumber\\
& \times \text{Pf}_{n_i=n_j=1}\left[\frac{1}{z_i-z_j}\right] e^{-\frac{1}{4} \sum_l |z_l|^2 n_l},
\end{align}
where the gauge factor has been set to one. It is not straightforward to take the continuum limit in this basis, since one has to define the limit of the Hilbert space on which the wave functions are defined. However, since the number of particles is conserved, we can rewrite the wave function in the basis spanned by the positions $w_1,\ldots,w_M$ of the particles. For $S=1/2$ there is at most one particle per site so the wave function can be simply expressed as
\begin{align}
\psi_{\frac{1}{2}}(w_1, \ldots, w_M)&\propto \prod_{i<j}(w_i-w_j)^{q} \text{Pf}\left[\frac{1}{w_i-w_j}\right] e^{-\frac{1}{4} \sum_l |w_l|^2},
\end{align}
where the $w_i$ are restricted to positions in the lattice. In the limit of infinitely many lattice sites per particle the lattice becomes a continuous plane and the positions $w_i$ become positions in the plane. This state then coincides with the Moore-Read state (\ref{MooreRead}).
The number of particles on the lattice is $M=\eta\frac{N}{q}=\frac{a}{2\pi}\frac{N}{q}$, so if the flux is $\frac{\text{Area}}{2\pi}$, then we can express $\frac{1}{q}=\frac{M}{\text{flux}}$. This explains that this quantity corresponds to the filling fraction in the continuum, defined as the number of particles per magnetic flux.

\subsection{Continuum limit of the $S=1$, $q=1$ state}
\label{subsec:limits1}

For $S=1$, $q=1$, the state also has the form (\ref{CFTcontinuum}), however since the $n_i$ can take the value $2$, it is not straightforward to take the continuum limit. We first have to write the wave function in the basis spanned by the position of the particles. For a basis element $\Ket{n_1,\ldots,n_N}$, let $w_r$, $r\in\{1,\ldots,M\}$ be the positions of the particles. Since we interpret the state $\Ket{2}$ as the presence of two particles, positions $z_i$ where $n_i=2$ are listed with two different indices $r$ in the set $\{w_r\}$. We now write the wave function in the basis given by the sets $\{w_r\}$.

As a starting point, observe that 
\begin{align}
\prod_{r=1}^M e^{-|w_r|^{2}/4}&=\prod_{i=1}^n e^{-n_i|z_{i}|^{2}/4}.\label{continuum4}
\end{align}

We will now prove that 
\begin{align}
\prod_{r<s}(w_r-w_s)^1 &\text{Pf}\left[\frac{1}{w_r-w_s}\right]=\nonumber\\
&\prod_{i<j}(z_i-z_j)^{n_i n_j}\text{Pf}_{n_i=n_j=1}\left[\frac{1}{z_i-z_j}\right].\label{continuum3}
\end{align}
Suppose first that only the first site is doubly occupied. Then,
\begin{align}
\prod_{r<s}(w_r-w_s)=(z_1-z_1)\prod_{1<i}(z_1-z_i)^{n_1 n_i}\prod_{2\leq i<j}(z_i-z_j)^{n_i n_j}.\label{continuum1}
\end{align}
Moreover,
\begin{align}
\text{Pf}\left[\frac{1}{w_r-w_s}\right]&=\frac{1}{z_1-z_1} \text{Pf}_{r,s\geq 2}\left[\frac{1}{w_r-w_s}\right],\\
&=\frac{1}{z_1-z_1} \text{Pf}_{n_i=n_j=1}\left[\frac{1}{z_i-z_j}\right],\label{continuum2}
\end{align}
where we have used in the last line that all other positions have at most one particle. Combining (\ref{continuum1}) and (\ref{continuum2}) we see that the $(z_1-z_1)$ term cancels, so that (\ref{continuum3}) holds. The same derivation with a recursion on the number of sites with two particles proves that (\ref{continuum3}) holds for all basis elements.

Putting together (\ref{continuum3}) and (\ref{continuum4}), the wave function can therefore be written as
\begin{align}
\label{wavecontinuum}
\psi_{1}^{q=1}(w_1, \ldots, w_M)&\propto \prod_{i<j}(w_i-w_j)^{1} \text{Pf}\left[\frac{1}{w_i-w_j}\right] e^{-\frac{1}{4} \sum_l |w_l|^2},
\end{align}
where the $w_r$ can be repeated twice to allow for states with double occupation, in which case this expression does not vanish because of a cancellation between the Jastrow factor and the Pfaffian. If we now take the continuum limit, the positions of the particles can be anywhere on the plane and this becomes the bosonic Moore-Read state (\ref{MooreRead}) at filling fraction 1, which also does not vanish when two particles are at the same site.

One may ask what happened to the doubly occupied site. In the continuum the ensemble of states with two particles at the same positions has measure zero compared to states with at most one particle at each position, therefore they are irrelevant and do not contribute to the wave function. Similarly, when we take the continuum limit of the lattice, states with two particles at the same site do not contribute to the wave function, as we will now explain in more details. Intuitively, this comes from the fact that there is a finite number $\sum_{i} n_i = M$ of particles, so configurations with at least one doubly occupied site are rare when the number of sites goes to infinity. More quantitatively, let us denote $P_m$ the number of basis elements $\Ket{n_1,\ldots,n_N}$ satisfying $\sum n_i = M$ with $m$ doubly occupied sites. $P_0=\binom{N}{M}$ is the number of basis elements with no doubly occupied sites and $P_m=\frac{N!}{m!(M-2m)!(N+m-M)!}$. In the continuum limit, $M$ and $m$ are fixed and $N$ goes to infinity, so $P_m/P_0\sim K/N^m$ where K is a constant and $\sum_{m=1}^{M/2} P_m/P_0 \rightarrow 0$. This shows that the number of basis elements with at least one doubly occupied site is small compared to the number of basis elements with no doubly occupied sites. Let us now observe that the wave function has to be normalized by a factor $Q=\sum_{n_1,\ldots,n_N} |\psi(n_1,\ldots,n_N)|^2$. We can decompose $Q$ between the basis of states with zero doubly occupied sites and the basis of states with at least one doubly occupied site :
\begin{align}
Q&=\sum_{n_1,\ldots,n_N} |\psi(n_1,\ldots,n_N)|^2,\\
&= \sum_{n_1,\ldots,n_N} \delta_{\sum n_i=M} |\psi(n_1,\ldots,n_N)|^2,\\
&= \sum_{\underset{\sum n_i=M}{\underset{\forall n_i\setminus n_i\neq 2}{n_1,\ldots,n_N}}} |\psi(n_1,\ldots,n_N)|^2+\sum_{\underset{\sum n_i=M}{\underset{\exists n_i\setminus n_i=2}{n_1,\ldots,n_N}}} |\psi(n_1,\ldots,n_N)|^2,
\end{align}
where the first sum contains $P_0$ elements and the second sum contains $\sum_{m=1}^{M/2}P_m\ll P_0$ elements. We now observe that elements appearing in these two sums are of the same order. Indeed, let us take a configuration with $M$ particles at positions $w_r$ such that 2 particles are at the same site ($w_1=w_2$), corresponding to an element $\psi(n_1,\ldots,n_N)$ in the second sum. Then by slightly moving one of the particles ($w_1=w_1+\delta$) we obtain a new configuration that is in the first sum, where all particles are at distinct positions. Since the number of particles is fixed, continuity of the coefficients of the wave function (\ref{wavecontinuum}) implies that the new element obtained is close to $\psi(n_1,\ldots,n_N)$. Therefore elements in the second sum are of the same order as elements in the first sum and since there are $\sum_{m=1}^{M/2}P_m\ll P_0$ elements in the second sum, in the continuum limit this sum does not contribute to the normalization factor Q. This shows that the contribution of configurations with two particles at one or more sites can be neglected in the continuum limit.

Note that the previous derivation also shows that the $(q=1,\eta=1)_1$ CFT state in the thermodynamic limit is equivalent to the spin 1 non-Abelian chiral spin liquid introduced in Ref.~\cite{Greiter2009}, but the two states are different on finite lattices.

\subsection{Continuum limit of the $S=1$, $q\geq 2$ state}

When $q\geq 2$ the state can be written as
\begin{align}
\psi_{1}^{q\geq 2}(n_1, \ldots, n_N)&\propto \prod_{i<j}(z_i-z_j)^{(q-1) n_i n_j} \psi_{1}^{q=1}(n_1, \ldots, n_N)
\end{align}
We have already derived the continuum limit of the state on the right at $q=1$ and the remaining factor can be expressed as 
\begin{align}
\prod_{i<j}(z_i-z_j)^{(q-1) n_i n_j}=\prod_{r<s}g(w_r,w_s)^{(q-1)},
\end{align}
where 
\begin{align}
g(w_r,w_s)=\left\{
\begin{array}{l}
  w_r-w_s \ \ \ \text{if $w_r \neq w_s$}  \\
  1\ \ \ \text{otherwise} 
\end{array}
\right.
\end{align}
In the continuum limit, the wave function can therefore be written as
\begin{align}
\psi_{1}^{q\geq 2}(w_1,\ldots,w_m)\propto&\prod_{r<s}g(w_r-w_s)^q \nonumber \\ & \times \text{Pf}\left[\frac{1}{w_r-w_s}\right] 
\prod_r e^{-|w_r|^{2}/4}.
\end{align}
When no two particles are at the same site, this wave function is the same as the $(q,\eta)_{1/2}$ CFT state in the continuum limit. As explained in Sec.~\ref{subsec:limits1}, configurations with two particles at the same site do not contribute to the wave function in the continuum limit, therefore the $(q,\eta)_{1}$ CFT states for $q\geq2$ have the same continuum limit as the $(q,\eta)_{1/2}$ CFT states, which is the Moore-Read state at filling fraction 1/q.

\subsection{One dimensional continuum limit}

So far we have focused on two dimensional states. In the one dimensional setting, when $z_j=e^{2\pi i j/N}$, the same results enable us to perform an interpolation between the lattice and the continuum. The only difference with the 2D case is that it is now possible to compute analytically
\begin{align}
f_N(z_l)\propto \xi_l z_l^\eta,
\end{align}
so the wave functions in the continuum can be expressed as 
\begin{align}
\psi_{S}(w_1, \ldots, w_M)&\propto \prod_{i<j}(w_i-w_j)^{q} \text{Pf}\left[\frac{1}{w_i-w_j}\right] \prod_j \xi_j,
\end{align}
which is a one-dimensional version of the Moore-Read state.

\section{Properties of the CFT states}
\label{section3}

In two dimensions, the Moore-Read states in the continuum are topological states which support non-abelian quasi-particle excitations. It is of high interest to check whether the lattice CFT states we have introduced share these properties. In one dimension, we expect that the CFT states display critical behaviour related to the conformal operators used to construct the wave function. In this section we focus on the states with $q=1$ and $q=2$, and numerically compute some of the properties of the states we have introduced.

\subsection{One dimensional critical states}

We now look at one-dimensional chains such that $z_j=e^{2\pi i j/N}$. Since we will find local Hamiltonians for the states $(q=1,\eta)_1$ and $(q=2,\eta)_{1/2}$ in one dimension in the lattice limit in Sec.~\ref{section5}, we focus on these states. First we compute the Renyi entropy $S^{(2)}_L = - \ln {\rm Tr} \;  \rho^2_L$, where $\rho_L$ is the density matrix of the CFT states restricted to a subsystem of size $L$ of the chain. This computation can be performed by using a Metropolis-Hastings algorithm with two independent spin chains \cite{cftimps,Hastings2010,Bajdich2008,Wimmer2012}.

\begin{figure}[h]\begin{center}
 % \subfigure[]{\includegraphics[scale=0.65]{images/decayexact.pdf}\label{sub2a}}
 \subfigure[]{\includegraphics[scale=0.4]{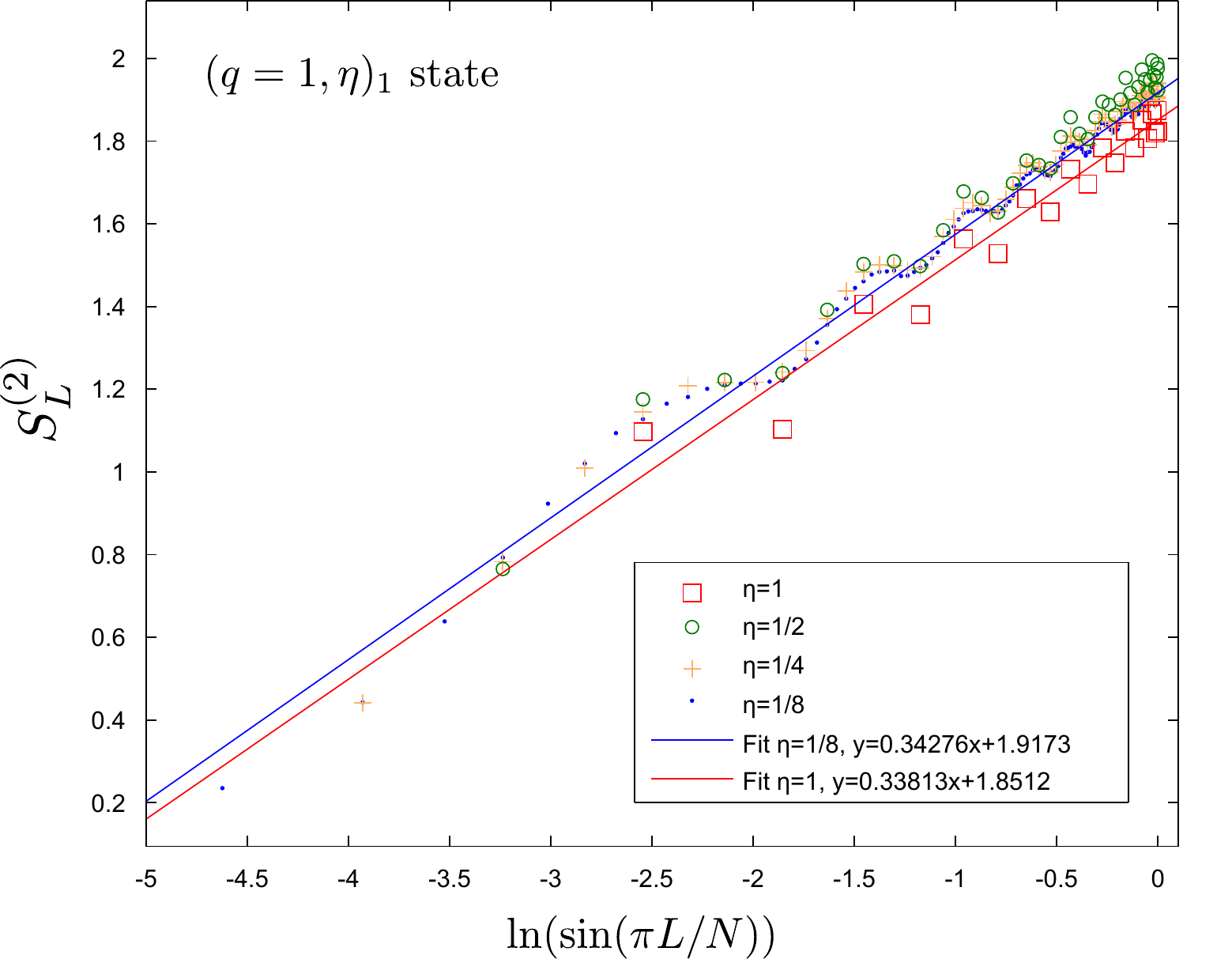}\label{fig:overlap1D}}
 \subfigure[]{\includegraphics[scale=0.4]{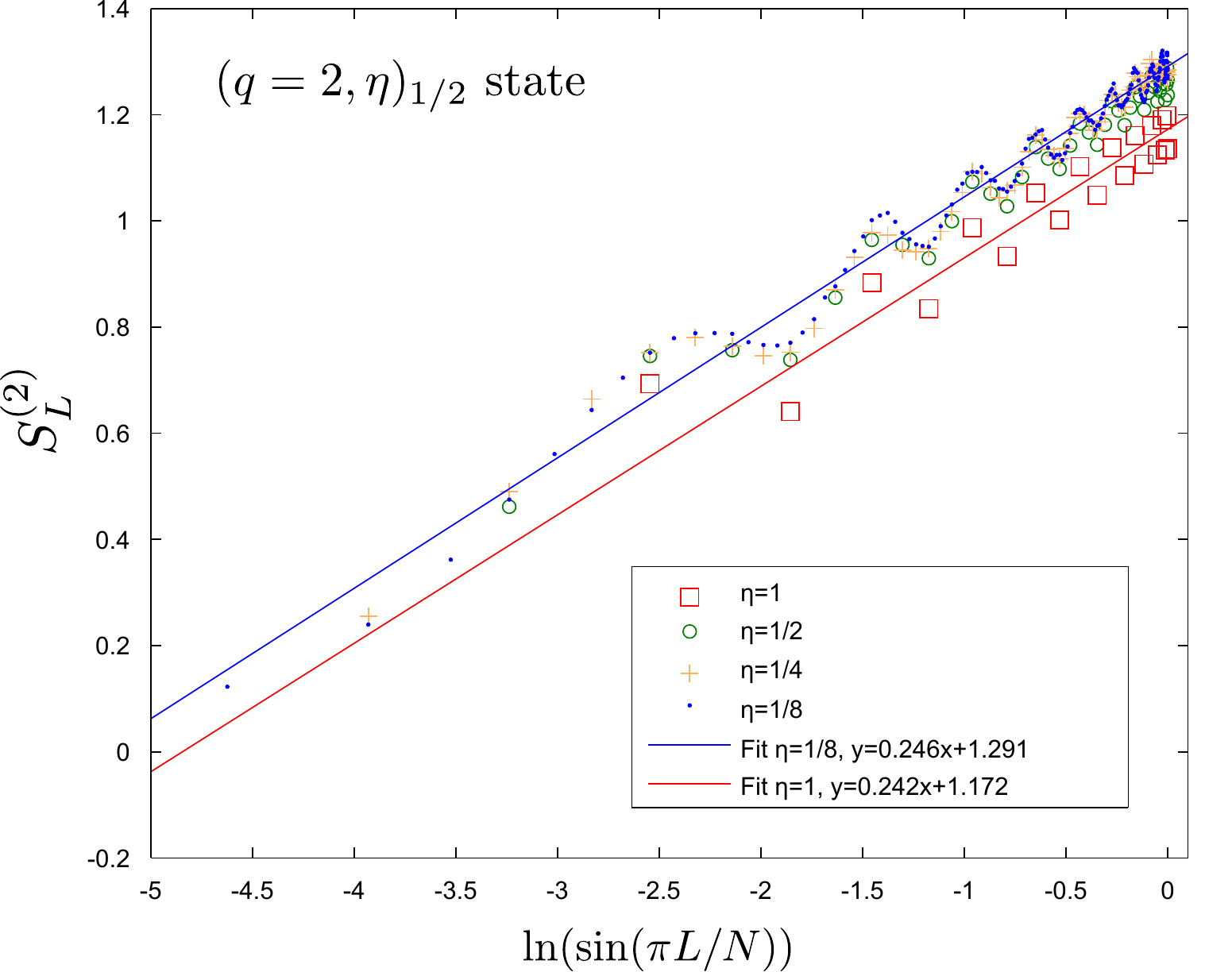}\label{fig:overlap1D2}}
\caption{\label{fig:ent1D} Renyi entropy $S_L^{(2)}$ of a subsystem of $L$ consecutive sites for the 1D $(q=1,\eta)_{1}$ CFT states (a) and $(q=2,\eta)_{1/2}$ CFT states (b) for different values of $\eta$. The number of particles $M=\eta N/q$ is fixed so the sizes of the chain are $N=40,80,160,320$ for $\eta=1,1/2,1/4,1/8$ respectively. The lines are linear fits of the points for $\eta=1/8$ (blue) and $\eta=1$ (red).}
\end{center}
\end{figure}

The results are shown in Fig.~\ref{fig:ent1D}. The entropy scales logarithmically with the size of the subchain for all values of $\eta$. Moreover, the scaling is approximately the same for different values of $\eta$ and fits of the form
\begin{align}
S^{(2)}_L=\frac{c}{4} \ln(\sin(\frac{\pi L}{N}))+b
\end{align} 
yield a value of the central charge c approximately equal to $1.36$ for the $(q=1,\eta)_{1}$ CFT states (the main source of errors is here the finite size of the lattice considered). This value is in agreement with the value of $1.395$ found for the state in the lattice limit in Ref.\cite{Nielsen2011}, where it was also shown that a value of $1.5$ for the central charge, as expected for the $SU(2)_2$ WZW model, could not be excluded. For the $(q=2,\eta)_{1/2}$ CFT states we find a value of $0.98$ which is compatible with a central charge equal to 1.

\begin{figure}[h]\begin{center}
 % \subfigure[]{\includegraphics[scale=0.65]{images/decayexact.pdf}\label{sub2a}}
 \subfigure[]{\includegraphics[scale=0.4]{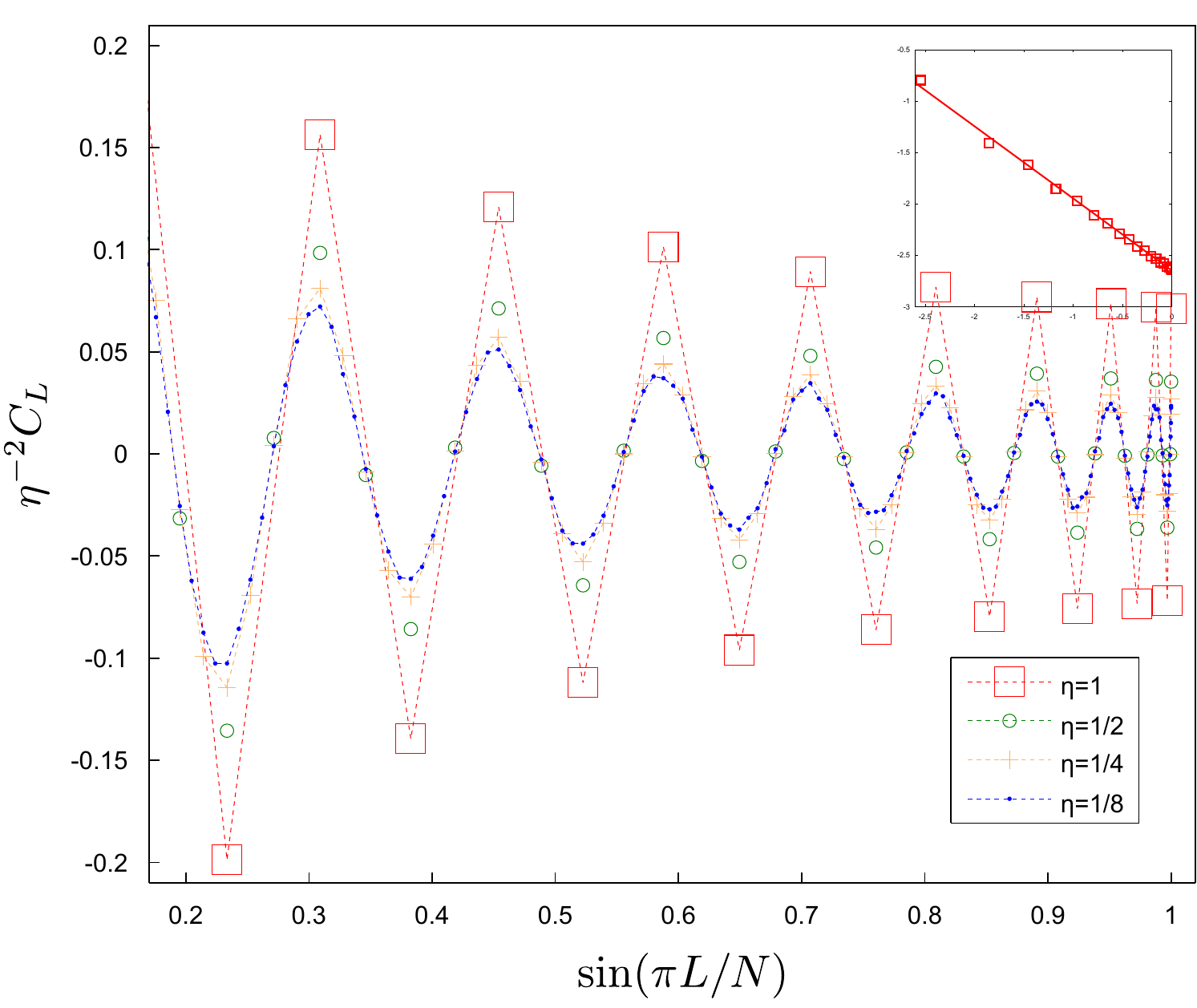}\label{fig:overlap1D}}
 \subfigure[]{\includegraphics[scale=0.4]{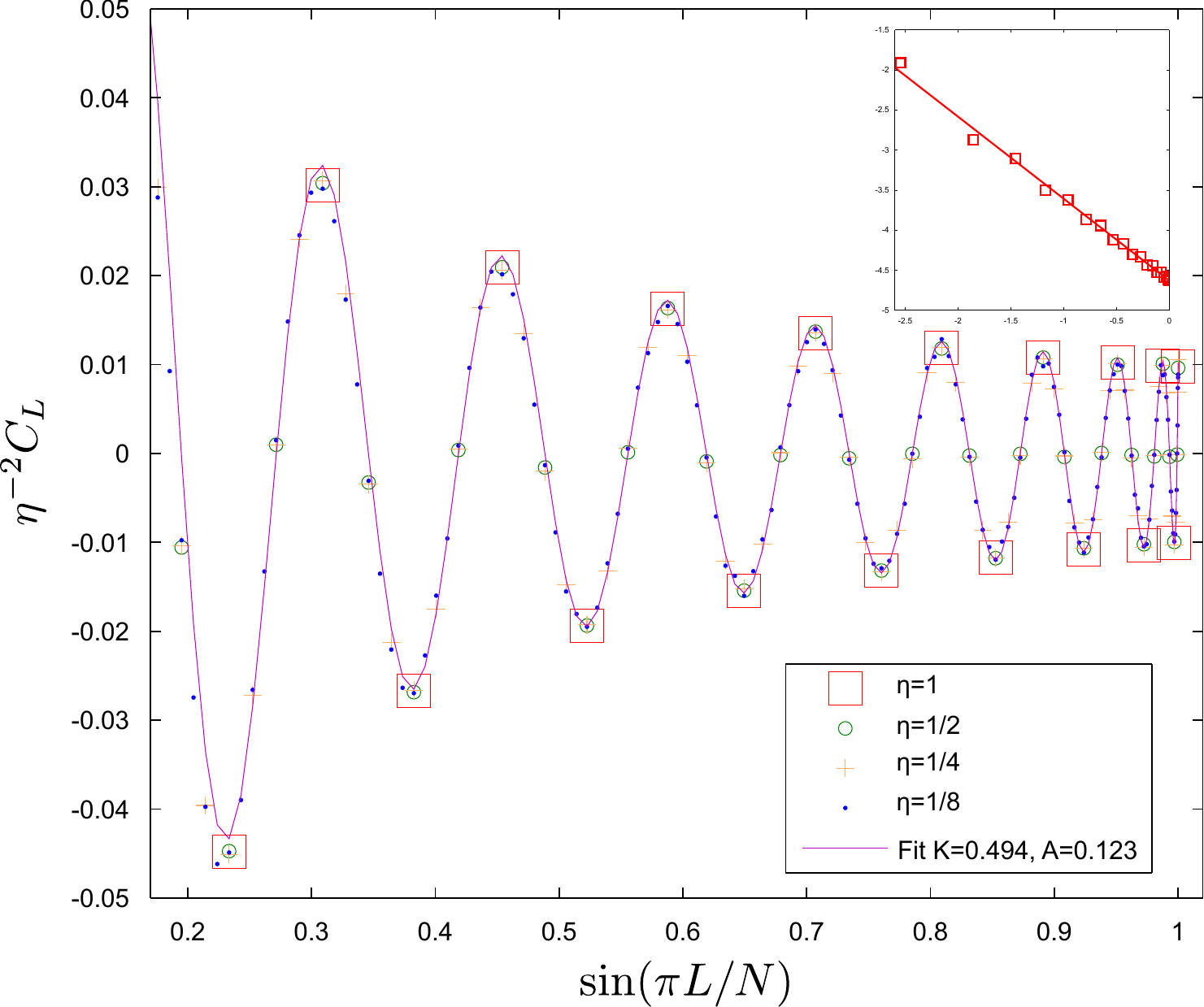}\label{fig:overlap1D2}}
\caption{\label{fig:corr1D} Rescaled correlation function $\eta^{-2} C_L$ as a function of the distance between the sites for the 1D $(q=1,\eta)_{1}$ CFT states (a) and $(q=2,\eta)_{1/2}$ CFT states (b) for different values of $\eta$. The number of particles $M=\eta N/q$ is fixed so the sizes of the chain are $N=40,80,160,320$ for $\eta=1,1/2,1/4,1/8$ respectively. The data for $\eta=1$ is shown in the insets in log-log scale, confirming the polynomial decay of correlations, and the line in the insets is a linear fit yielding critical exponents $0.70$ ($(q=1,\eta=1)_{1}$ state) and $1.02$ ($(q=2,\eta=1)_{1/2}$ state). In (b), the line is a fit of the form Eq. (\ref{corrTLL}) with parameters K=0.494 and A=0.123.}
\end{center}
\end{figure}

Another quantity that can be computed using Monte-Carlo techniques is the particle-particle correlation function $C_L=\Braket{n_1 n_L}-\Braket{n_1}\Braket{n_L}$. Results in Fig.~\ref{fig:corr1D} confirm that the states are critical since the correlation functions decay polynomially with the distance $L$. For the $(q=1,\eta=1)_{1}$ CFT state, the critical exponent is found to be $0.70$, which is in agreement with the value of $0.69$ found in Ref.\cite{Nielsen2011}, where it was observed that such a value can be influenced by a multiplicative logarithmic correction which could explain the difference with the expected value of $0.75$ \cite{Affleck1986,Affleck1989,Narayan2004}. Moreover for the $(q=2,\eta)_{1/2}$ states at different values of $\eta$, the correlations are very close once rescaled by a factor of $1/\eta^2$, which confirms that properties of the state do not change along the interpolation. For a Tomonaga-Luttinger liquid, the expected behaviour of the particle-particle correlation function is \cite{Cabra2004}
\begin{align}
\label{corrTLL}
C_L=\frac{A\cos(2L k_F)}{\left|\sin(\frac{\pi L}{N})\frac{N}{\pi}\right|^{2K}}+\frac{K}{2\pi^2\left|\sin(\frac{\pi L}{N})\frac{N}{\pi}\right|^2},
\end{align} 
where $K$ is the Luttinger parameter, $k_F=\eta \pi/q$ is the Fermi momentum and A is a non-universal constant. For the $(q=2,\eta)_{1/2}$ state we find a good agreement of this formula for $K=0.494$, $A=0.123$. This suggest that this state in one dimension is well described by a Tomonaga-Luttinger liquid with central charge $c=1$ and Luttinger parameter $K=0.5$, which corresponds to the properties of a free-boson CFT with radius $\sqrt{2}$, as was the case for the corresponding one-dimensional Laughlin state\cite{Tu2014}.

\subsection{Two dimensional topological states}

In the continuum, the Moore-Read state at filling fraction $1/q$ has a topological entanglement entropy of \cite{Zozulya2007}
\begin{align}
\gamma_{0}(q)=\frac{1}{2}\ln (4q).
\end{align}

\begin{figure}[htb]
\centering
\def\svgwidth{7.6cm}
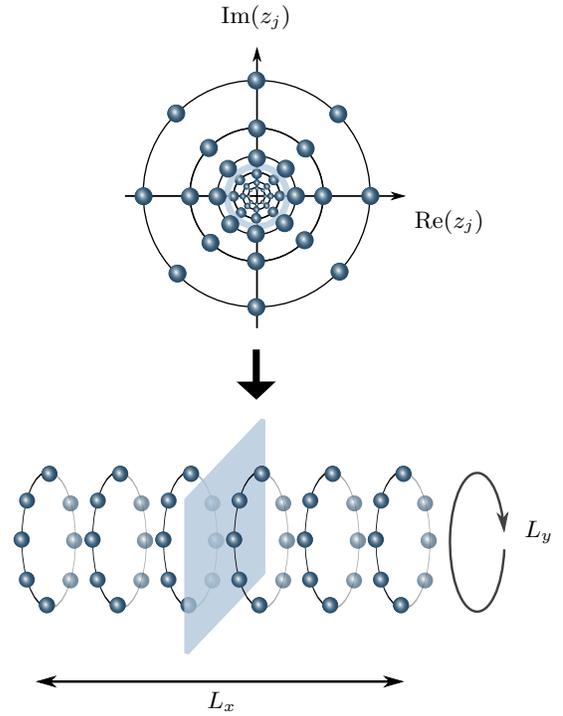
\caption{\label{fig:cylinder}
  The mapping from a lattice on the complex plane to a cylinder. To compute the topological entanglement entropy of the state, the cylinder is cut into two halves and the Renyi entropy of the first half is computed using a Metropolis-Hastings algorithm. The topological entanglement entropy is then extracted by varying the size $L_y$ of the cylinder (\ref{TEE}).
}
\end{figure}

\begin{figure}[htb]\begin{center}
\subfigure{\includegraphics[scale=0.35]{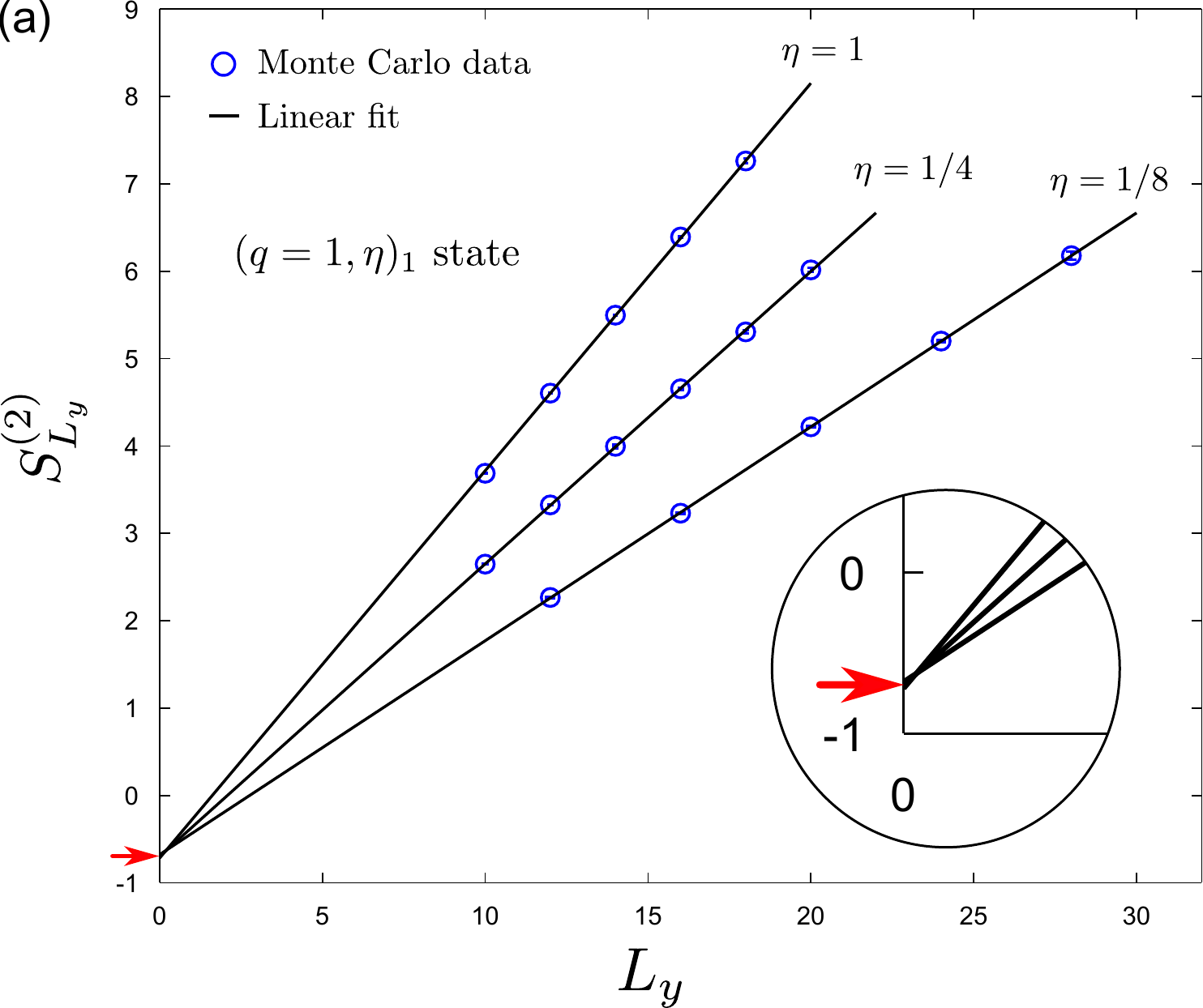}}  
\subfigure{\includegraphics[scale=0.35]{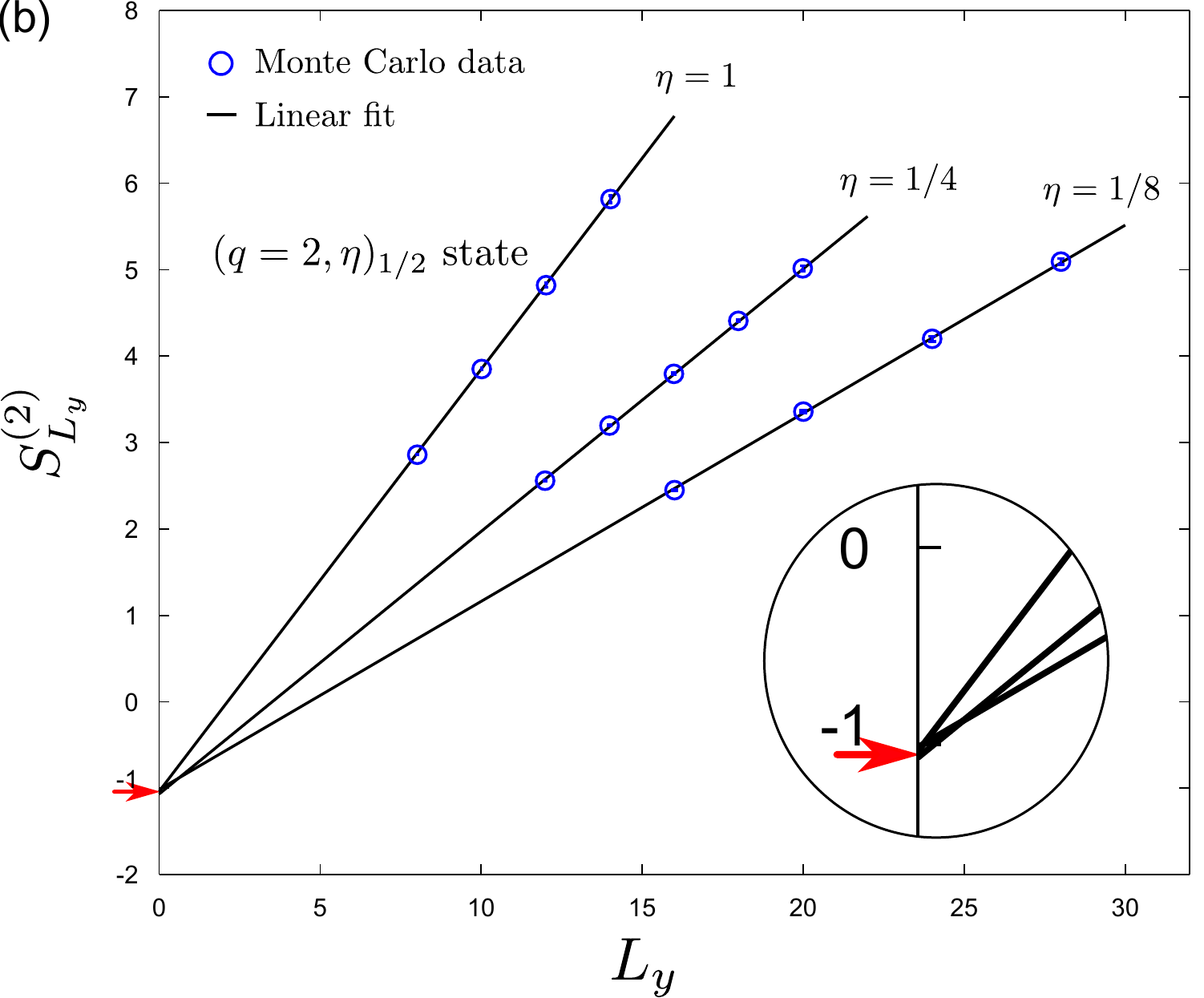}}
\subfigure{\includegraphics[scale=0.35]{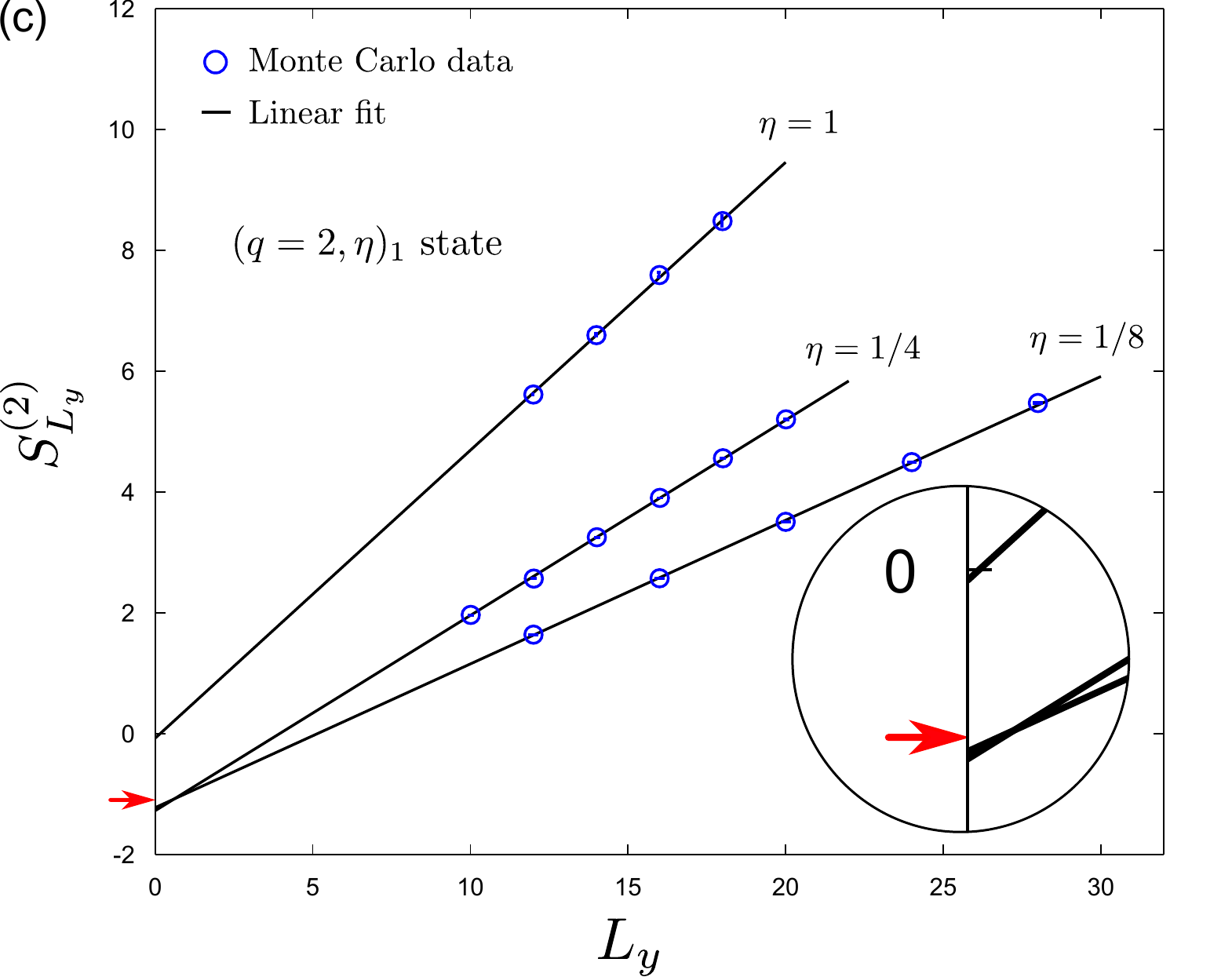}}
\caption{\label{fig:TEE11}
 Linear behaviour of the Renyi entropy with the size of the cut $L_y$ for the $(q=1,\eta)_{1}$ (a), $(q=2,\eta)_{1/2}$ (b) and $(q=2,\eta)_{1}$ (c) CFT states on a $L_x\times L_y$ lattice. The topological entanglement entropy of the Moore-Read states at filling 1 (a), $\gamma_{0}(1)\approx 0.69$, and at filling $1/2$ (b,c), $\gamma_{0}(2)\approx 1.04$, are indicated with a red arrow. The values of $\eta$ are 1, $1/4$ and $1/8$ and the corresponding sizes $L_x$ are respectively $12$, $16$ and $16$. The insets are enlarged views confirming that the topological entanglement entropy stays the same when $\eta$ is varied and that its value corresponds to $\gamma_{0}(1)$ (resp. $\gamma_{0}(1/2)$) in the first two cases, while the topological entanglement entropy of the $(q=2,\eta)_{1}$ CFT state is close to zero in the lattice limit and close to $\gamma_{0}(1/2)$ in the continuum limit.
}
\end{center}
\end{figure}

To compute the topological entanglement entropy (TEE) of the CFT states, we map a $L_x\times L_y$ square lattice on the cylinder to the complex plane by choosing the positions of the lattice sites to be $z_j=e^{2\pi((x_j-L_x-1/2)+iy_j)/L_y}$, where $x_j\in\{1,\ldots,L_x\}$, $y_j\in\{1,\ldots,L_y\}$ (Fig. \ref{fig:cylinder}). We then cut the cylinder in two halves and compute the Renyi entropy of the first half. The size along the cut is $L_y$ and we use the behaviour of the entanglement entropy \cite{Kitaev2006,Levin2006}
\begin{align}
\label{TEE}
S^{(2)}_{L_y}=\alpha L_y - \gamma
\end{align}
to extract the topological entanglement entropy $\gamma$ (Fig. \ref{fig:TEE11}). The results for the states $(q=1,\eta)_{1}$ (resp. $(q=2,\eta)_{1/2}$) are in agreement with the topological entanglement entropy of a Moore-Read state at filling fraction 1 (resp. $1/2$), $\gamma_{0}(1)=\frac{1}{2}\ln(4)\approx 0.69$ (resp. $\gamma_{0}(2)=\frac{1}{2}\ln(8)\approx 1.04$). Moreover the value of the TEE does not change with $\eta$, so topological properties of the states remain the same along the interpolation between the continuum and the lattice limit.

We observe however that the state $(q=2,\eta)_{1}$ has a TEE close to zero and different than $\gamma_{0}(2)$. The TEE does not stay constant when $\eta$ is changed, which is compatible with the expectation that its value is $\gamma_{0}(2)$ in the continuum limit :  there must be a phase transition along the interpolation between the continuum and the lattice. The states with $S=1$, $q\geq 2$ can therefore define distinct lattice states from the states with $S=1/2$, $q\geq 2$, while having the same continuum limit.

\section{Parent Hamiltonians}
\label{section4}

So far we have considered wave functions for lattice states. It is also relevant to ask whether these states are ground states of some Hamiltonians and whether these Hamiltonians can be realized in nature or implemented in experiments. We turn now to the construction of parent Hamiltonians for which the CFT states are ground states. The CFT states are constructed from correlators of conformal fields. As has been shown in Ref. \cite{Nielsen2011}, this enables the construction of parent Hamiltonians from null fields of the considered CFT. Null fields are fields such that when inserted in a correlator of primary fields, the evaluation of the correlator gives zero. The procedure is as follows : 
\begin{enumerate}
\item Find null fields $\chi^a(z_i)$ labelled by $a$ and acting at position $z_i$ of the considered CFT.
\item The vacuum expectation value of a product of primary chiral conformal fields is zero if one of the fields is a null vector, therefore we get equations
\begin{align}
\Braket{\mathcal{V}_{n_1}(z_1) \dots \chi^a(z_i) \dots \mathcal{V}_{n_N}(z_N)}=0.
\end{align}
\item These equations are rewritten in the form $\Lambda^a_i \Ket{\psi}=0$, where $\Lambda^a_i$ are operators acting on the degrees of freedom of the wave function.
\item $\sum_{a,i} \Lambda^{a\dagger}_i\Lambda^a_i$ is then a positive semi-definite Hermitian operator annihilating the wave function.
\end{enumerate}
In this section we apply this procedure to construct parent Hamiltonians for the CFT states in the lattice limit. In the rest of this work the phase factors are fixed to $\xi_l=1$.

\subsection{Parent Hamiltonians for the $SU(2)_2$ $(q=1,\eta=1)_{1}$ CFT state}
\label{parentq1}
The special case of the $(q=1,\eta=1)_{1}$ state has a wave function constructed from the spin 1 primary fields of the $SU(2)_2$ Wess-Zumino-Witten conformal field theory \cite{Ardonne2010,Nielsen2011} and has already been considered partially in Ref.~\cite{Nielsen2011}, where, however, the focus was on 1D systems. This $SU(2)_2$ symmetry can be used to construct parent Hamiltonians invariant under $SU(2)$ transformations. 

As shown in Ref.~\cite{Nielsen2011}, the null fields in this case can be parametrized by
\begin{align}
\chi^a(z_i)=\left(K^{(i)}\right)^a_b (J_{-1}^b \varphi_n) (z_i),
\end{align}
where repeated indices are summed over, $\varphi_n$ is a chiral spin 1 primary field, $J_{-1}^b$ are the $-1$ modes of the $SU(2)_2$ current operators and 
\begin{align}
\left(K^{(i)}\right)^a_b=\frac{2}{3} \delta_{ab} - \frac{5}{12} i \epsilon_{abc} t_i^c-\frac{1}{12} (t_i^a t_i^b+t_i^b t_i^a),
\end{align}
where $\epsilon_{abc}$ is the Levi-Civita symbol and $t_i^a$ are the spin 1 operators acting on site i. These operators can be written in the spin basis at site i as
\begin{align}
t_i^x=\frac{\hbar}{\sqrt{2}}\begin{pmatrix}
0&1&0\\
1&0&1\\
0&1&0
\end{pmatrix},\ \ \ \
t_i^y=&\frac{\hbar}{\sqrt{2}}\begin{pmatrix}
0&-i&0\\
i&0&-i\\
0&i&0
\end{pmatrix},\nonumber\\
t_i^z=\hbar\begin{pmatrix}
1&0&0\\
0&0&0\\
0&0&-1
\end{pmatrix}&.
\end{align}

We then exploit that the correlator with a null field inserted is zero :
\begin{align}
\Braket{\varphi_{n_1}(z_1) \dots \chi^a(z_i) \dots \varphi_{n_N}(z_N)}=0.\label{hamiltoSU21}
\end{align}
The Ward identity enables to transfer the action of a current operator to fields at other positions :
\begin{align}
&\Braket{\varphi_{n_1}(z_1) \dots (J_{-1}^b \varphi_{n_i}) (z_i) \dots \varphi_{n_N}(z_N)}=\nonumber\\
&\sum_{j(\neq i)}^N \frac{t_j^a}{z_1-z_j}\Braket{\varphi_{n_1}(z_1) \dots \varphi_{n_i} (z_i) \dots \varphi_{n_N}(z_N)}.\label{hamiltoSU22}
\end{align}
Using (\ref{hamiltoSU22}) it is possible to rewrite (\ref{hamiltoSU21}) as \cite{Nielsen2011}
\begin{align}
\Lambda_i^{a}\Ket{\psi}&=0,
\end{align}where
\begin{align}
\Lambda_i^{a}&=\sum_{j(\neq i)}^N \left(K^{(i)}\right)^a_b \omega_{ij} t_j^b,\\
&=\sum_{j(\neq i)}^N  \omega_{ij} \left[\frac{2}{3} t_j^a - \frac{5}{12} i \epsilon_{abc} t_j^b t_i^c-\frac{1}{12} (t_i^a t_i^b+t_i^b t_i^a)t_j^b\right],
\end{align}
where we have defined $\omega_{ij}=\frac{z_i+z_j}{z_i-z_j}$. This can be used to construct a parent Hamiltonian $H=\sum_{a,i} \Lambda^{a\dagger}_i\Lambda^a_i$, which gives
\begin{align}
H=\frac{4}{3}\sum_{i\neq j}^N \omega_{ij}^*\omega_{ij}+\frac{1}{3}\sum_{i\neq j}^N \left(\omega_{ij}^*\omega_{ij}+2\sum_{k(\neq i, j)}^N \omega_{ki}^*\omega_{kj}\right) t_i^a t_j^a \nonumber\\
-\frac{1}{6} \sum_{i\neq j}^N \omega_{ij}^*\omega_{ij} (t_i^a t_j^a)^2 + \sum_{i\neq j \neq k} \left(\frac{1}{3}\omega_{ik}^*\omega_{ij} - \frac{1}{2}\omega_{ik}\omega_{ij}^*\right) t_i^a t_j^a t_i^b t_k^b \label{hamiltoSU2}.
\end{align}
This Hamiltonian is $SU(2)$ invariant and numerical diagonalization on small systems confirm that it has the $(q=1,\eta=1)_{1}$ CFT state as a unique ground state. This Hamiltonian is similar to the one obtained in Ref.~\cite{Greiter2014} for the spin 1 non-Abelian chiral spin liquid state introduced in Ref.~\cite{Greiter2009}. However the Hamiltonian we just constructed is valid for any choice of lattice and not only in the thermodynamic limit as is the case in Ref.~\cite{Greiter2014}.

\subsection{Parent Hamiltonians for the $(q,\eta=1)_{1}$, $q\geq2$ CFT states}

For $q\geq 2$, the $(q,\eta=1)_{1}$ CFT states do not display an $SU(2)$ symmetry. However it is still possible to find null fields and construct operators annihilating the wave function. Let us define the operators $G^{\pm }(z)=\chi(z) e^{\pm i\sqrt{q}\phi (z)}$, $J(z)=\frac{i}{\sqrt{q}}\partial \phi (z)$. We use the following $q+1$ null fields :

\begin{align}
\underset{p=0,1,\ldots,q-2}{\chi^{p}(w)} &=\oint_{w}\frac{dz}{2\pi i}\frac{1}{(z-w)^{p}} G^{+}(z) V_{1}(w), \label{eq41}\\                        
\chi^{q-1}(w) &=\oint_{w}\frac{dz}{2\pi i}\left[\frac{1}{(z-w)^{q-1}} G^{+}(z) V_{1}(w)\right.\nonumber\\
&\ \ \ \ \ \ \ \ \ \ \ \ \ \ \ \ \left.-\frac{1}{(z-w)}V_{2}(w)\right],\\
\chi^{q}(w) &=\oint_{w}\frac{dz}{2\pi i}\frac{1}{z-w} \left[\frac{1}{(z-w)^{q-1}}G^{+}(z)V_{1}(w)\right] \nonumber \\
&\ \ \ - \oint_{w}\frac{dz}{2\pi i}\frac{1}{z-w} qJ(z)V_{2}(w),
\end{align}
where Eq.~(\ref{eq41}) gives $q-1$ null fields $\chi^{p}(w)$ with $p$ running from $0$ to $q-2$. Let us define the operators $d^\dagger$, $d$ to be creation and annihilation operators (bosonic for $q$ odd, fermionic for $q$ even) acting between states $\Ket{0}$ and $\Ket{1}$, and $d'^\dagger$, $d'$ to be creation and annihilation operators acting between states $\Ket{1}$ and $\Ket{2}$. The number of particles at site $i$ is thus $n_i=n_i^{(1)}+2n_i^{(2)}$, where $n_i^{(1)}=d^\dagger d$ and $n_i^{(2)}=d'^\dagger d'$. Using the previous null fields, we derive the following operators annihilating the wave function (see Appendix \ref{appendixA} for the complete derivation).

\begin{align}
 \Lambda^{0} &=\sum_i d_i,\\
 \underset{p=1,\ldots,q-2}{\Lambda_i^{p}} &=\sum_{j(\neq i)} \frac{1}{(z_i-z_j)^{p}} d_j d_i'^\dagger,\\
 \Lambda_i^{q-1} &=\sum_{j(\neq i)} \frac{1}{(z_i-z_j)^{q-1}} d_j d_i'^\dagger+n_i^{(2)},\\
 \Lambda_i^{q} &=\sum_{j(\neq i)}\frac{1}{(z_i-z_j)^q}d_j d_i'^\dagger-\sum_{j(\neq i)}\frac{qn_j-1}{z_i-z_j}n_i^{(2)}.
\end{align}

This leads to a three-body Hamiltonian annihilating the wave function
\begin{align}
H=\sum_{i=1}^N \sum_{a=0}^q \Lambda_i^{a\dagger}\Lambda_i^{a}+\left(\sum_{i} n_i-\frac{N}{q}\right)^2,
\end{align}
where the last term fixes the number of particles.

This Hamiltonian annihilates the wave function $(q,\eta=1)_1$, however we find numerically that the ground space of this Hamiltonian is degenerate when $q \geq 3$ and that the degeneracy does not depend on the number of sites. Other simple null fields of the theory constructed with the same current operators do not lead to operators acting on the wave function that would reduce this degeneracy.

\subsection{Parent Hamiltonians for the $(q,\eta=1)_{1/2}$, $q\geq 2$ CFT states}

We can now use the previous results to construct parent Hamiltonians for the $(q,\eta=1)_{1/2}$, $q\geq 2$ CFT states. They are projections of the $(q,\eta=1)_{1}$ CFT states in the subspace allowing only for single occupation at each site, that we denote $\mathcal{H}^{1}$. Let us also define the Hilbert space $\mathcal{H}^{2}$ spanned by basis elements containing at least one site with two particles. We will now project the operators annihilating the $(q,\eta=1)_{1}$ CFT states onto $\mathcal{H}^{1}$ in order to get operators annihilating the $(q,\eta=1)_{1/2}$ CFT states (see Appendix \ref{appendixB} for the detailed derivation). We start by multiplying the operators $\Lambda_i^{a}$ obtained previously on the left by $d_i'$ : these operators continue to annihilate the $(q,\eta=1)_{1}$ CFT states. Since $d_i'd_i'^\dagger=n_i^{(1)}$, we get operators $\Lambda_i^{0}, d_i'\Lambda_i^{1}, \ldots, d_i'\Lambda_i^{q-2}$ that act on $\mathcal{H}^{1}$ and are zero on $\mathcal{H}^{2}$, so they also annihilate the $(q,\eta=1)_{1/2}$ CFT states. We can then use the fact that $d_i'\Lambda_i^{q-1}=\sum_{j(\neq i)} \frac{1}{(z_i-z_j)^{q-1}} d_j n_i^{(1)}+d_i'$ annihilates the $(q,\eta=1)_{1}$ wave function to replace the operator $d_i'$ in $d_i'\Lambda_i^{q}$ by $-\sum_{h(\neq i)} \frac{1}{(z_i-z_h)^{q-1}} d_h n_i^{(1)}$. The resulting operator $d_i'\Lambda_i^{q}$ then acts separately on $\mathcal{H}^{1}$ and $\mathcal{H}^{2}$. We finally get operators annihilating the $(q,\eta=1)_{1/2}$ wave function :

 \begin{align}
 \Lambda^{'0} &=\sum_i d_i,\\
  \underset{p=1,\ldots,q-2}{\Lambda_i^{'p}} &=\sum_{j(\neq i)} \frac{1}{(z_i-z_j)^{p}} d_j n_i^{(1)},\\
 \Lambda_i^{'q-1} &=\sum_{j(\neq i)}\frac{1}{(z_i-z_j)^q}d_j n_i^{(1)}\nonumber\\
    &\ \ \ +\sum_{j(\neq i)}\sum_{h(\neq i)} \frac{1}{(z_i-z_h)^{q-1}}\frac{qn_j^{(1)}-1}{z_i-z_j} d_h n_i^{(1)}.
 \end{align}

This leads to a five-body Hamiltonian
\begin{align}
H=\sum_{i=1}^N \sum_{a=0}^{q-1} \Lambda_i^{'a\dagger}\Lambda_i^{'a}+\left(\sum_{i}n_i^{(1)}-\frac{N}{q}\right)^2.
\end{align}

As in the previous case, this parent Hamiltonian has a single ground state only when $q=2$ and this ground state is the $(q=2,\eta=1)_{1/2}$ CFT state.

\section{Local Hamiltonians}
\label{section5}

The parent Hamiltonians we have derived involve three or five-body interactions between all sites on the lattice. These Hamiltonians would therefore be very difficult to implement in experiments. However in some cases it has turned out that states constructed from correlators of conformal fields had very high overlaps with ground states of local Hamiltonians \cite{cftimps,Tu2014328,Nielsen2013,Glasser2014}. This has lead to a protocol to implement one of these states in experiments \cite{Nielsen2013,Nielsen2014imp}. In this section we show that there is a local Hamiltonian for which the ground state is close to the $(q=1,\eta=1)_1$ CFT state in one and in two dimensions and that this result is also true for the $(q=2,\eta=1)_{1/2}$ CFT state in one dimension.

\subsection{Local Hamiltonians for the $(q=1,\eta=1)_{1}$ CFT state}

In one dimension the case of the $(q=1,\eta=1)_{1}$ CFT state was studied in Ref.~\cite{Nielsen2011}. It was shown that this state has a high overlap with the ground state of the bilinear-biquadratic spin 1 Hamiltonian
\begin{align}
H^{(1)}_{1D}=\sum_{i=1}^N \left[\cos{(\beta)} t_i^a t_{i+1}^a + \sin{(\beta)} \left(t_i^a t_{i+1}^a\right)^2\right],
\end{align}
with periodic boundary conditions, when $\beta=−0.3213$. Note that this Hamiltonian includes the 2-body terms present in the parent Hamiltonian (\ref{hamiltoSU2}).\\

\begin{table}[htb]
  \caption{\label{table:LocalHamilto}
  Terms in the Hamiltonian $H_{2D}$ and coefficients obtained after numerical optimization on a $4\times 4$ lattice.}
\centering
\renewcommand{\arraystretch}{2}
\begin{tabular}{c|c|l}
\hline
\hline
Operator & Configuration & Coefficient\\
\hline
\multirow{2}{*}{$t_i^a t_j^a$}&\multirow{2}[2]{*}[0mm]{\includegraphics[scale=0.5]{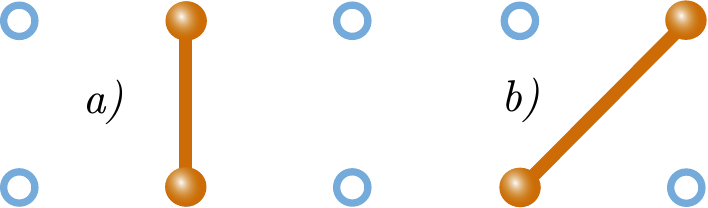}} & $a)$  1\\
& &  $b)$  0.6227\\
\hline
\multirow{2}{*}{$(t_i^a t_j^a)^2$}&  \multirow{2}{*}{\includegraphics[scale=0.5]{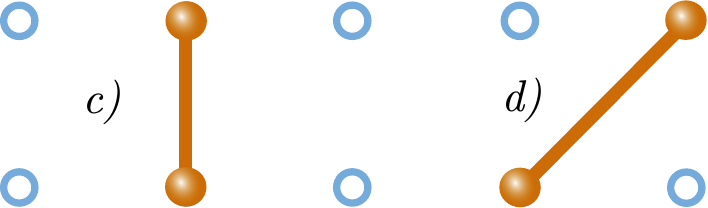}} & $c)$  -0.1762\\
& &  $d)$  0.3226\\
\hline
\multirow{2}{*}{$t_i^a t_j^a t_i^b t_k^b$}&  \multirow{2}{*}[-1mm]{\includegraphics[scale=0.5]{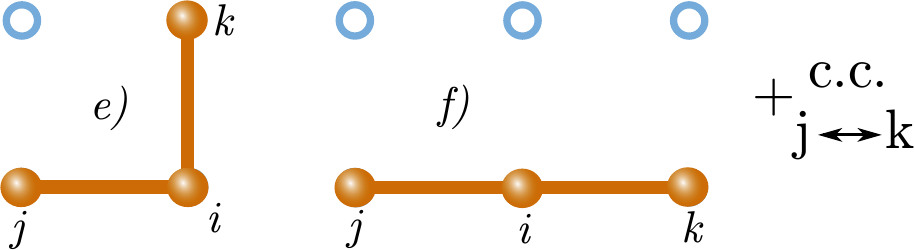}} & $e)$  0.4637$i$\\
& & $f)$  0.0208\\
\hline
\hline
\end{tabular}
\end{table}

We now study the two dimensional case and build a local Hamiltonian from the parent Hamiltonian (\ref{hamiltoSU2}). The operators $\Lambda_i^a$ contain 2-body interactions between sites $i$ and $j$. We cut these operators by keeping only terms for which the sites $i$ and $j$ are nearest-neighbours on the square lattice. This leads to a local Hamiltonian with three-body interactions. In addition to these terms, we include the two-body interactions between next-nearest neighbours present in the parent Hamiltonian. All six terms included in our trial Hamiltonian, that we denote $H_{2D}$, are shown in Table \ref{table:LocalHamilto}. Note that the coefficients of these terms in the exact parent Hamiltonian are not position-independent. In our local Hamiltonian, however, we choose them to be position-independent and invariant under rotations. 

By exact diagonalization and optimization on these coefficients, we find that there is a local Hamiltonian for which the overlap $|\Braket{\psi_H | \psi_{\text{CFT}}}|$ between the ground state and the $(q=1,\eta=1)_{1}$ CFT state on a $4\times4$ square lattice on the plane is $97.36\%$. Considering the size of the Hilbert space $3^{16}\approx 4\times 10^7$, this overlap is very high. Note that with the same parameters, the overlap is also above $98\%$ on a $4\times 3$ or on a $4 \times 2$ lattice. On a cylinder geometry, i.e. periodic boundary conditions in one direction, the overlap on a $4\times4$ square lattice is $97.21\%$. 

\begin{figure}[ht]
\centering
\begin{scriptsize}
\def\svgwidth{6.7cm}
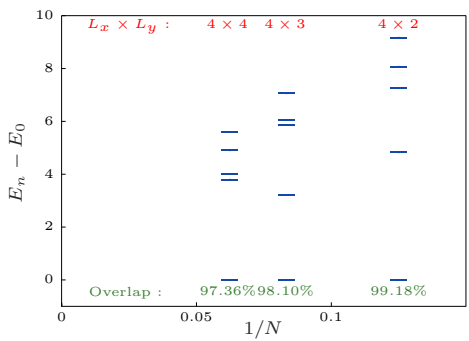
\end{scriptsize}
\caption{\label{fig:spectrum}
Energy difference to the ground state energy for the first excited states of the Hamiltonian $H_{2D}$ for different sizes of lattices. The overlap between the ground state of this Hamiltonian and the $(q=1,\eta=1)_{1}$ CFT state is indicated below each ground state. }
\end{figure}

Compared to the local Hamiltonian found in Ref. ~\cite{Greiter2009}, which is for a state that is equivalent to the $(q=1,\eta=1)_{1}$ CFT state in the thermodynamic limit, but different on finite lattices, the Hamiltonian we find has less free parameters (5 instead of 11) to fine-tune, which might make it easier to implement. Moreover, the very good scaling with lattice sizes lets us expect that a good agreement will persist on larger lattices. In Fig. \ref{fig:spectrum}, we show the low-energy spectrum of this local Hamiltonian. This figure is compatible with having a gap in the thermodynamic limit, but the limitations on the system sizes that we can consider prevent us from making a reliable extrapolation.

\subsection{Local Hamiltonian for the 1D $(q=2,\eta=1)_{1/2}$ CFT state}

The state $(q=2,\eta=1)_{1/2}$ CFT state has a five-body parent Hamiltonian. Let us consider the one dimensional case. If we cut the $\Lambda_i^a$ operators by keeping only terms for which the sites $i$ and $j$ are nearest-neighbours, we get a local Hamiltonian with several terms. We find however numerically that even a smaller number of terms is already sufficient to get a good overlap. Specifically, we choose to keep only the simplest two-body and the simplest three-body terms, to obtain a local Hamiltonian with periodic boundary conditions
\begin{align}
H^{(2)}_{1D}=\sum_{i=1}^N \left( n_i n_{i+1} n_{i+2} + \kappa  d_i^\dagger d_{i+1} \right) + c.c.,
\end{align}
where $d_i^\dagger$, $d_i$ are fermionic creation and annihilation operators at site $i$ and $n_i=d_i^\dagger d_i$. For $\kappa=0.274+0.052 i$, we find that the overlap between the ground state of $H^{(2)}_{1D}$ and the $(q=2,\eta=1)_{1/2}$ CFT state is $97.71\%$ for a chain with 20 spins (Fig. \ref{fig:overlap1D}).\\

\begin{figure}[h]\begin{center}
 % \subfigure[]{\includegraphics[scale=0.65]{images/decayexact.pdf}\label{sub2a}}
 \subfigure[]{\includegraphics[scale=0.4]{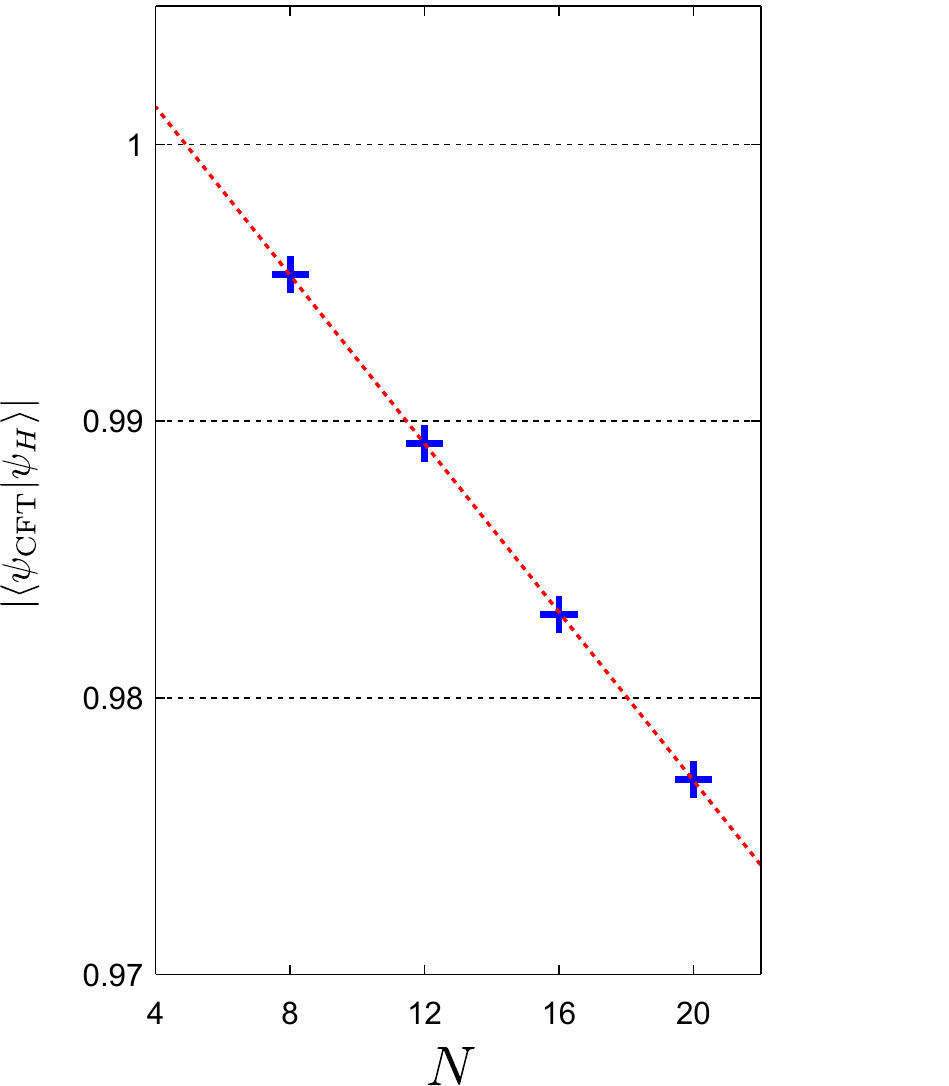}\label{fig:overlap1D}}
 \subfigure[]{\includegraphics[scale=0.4]{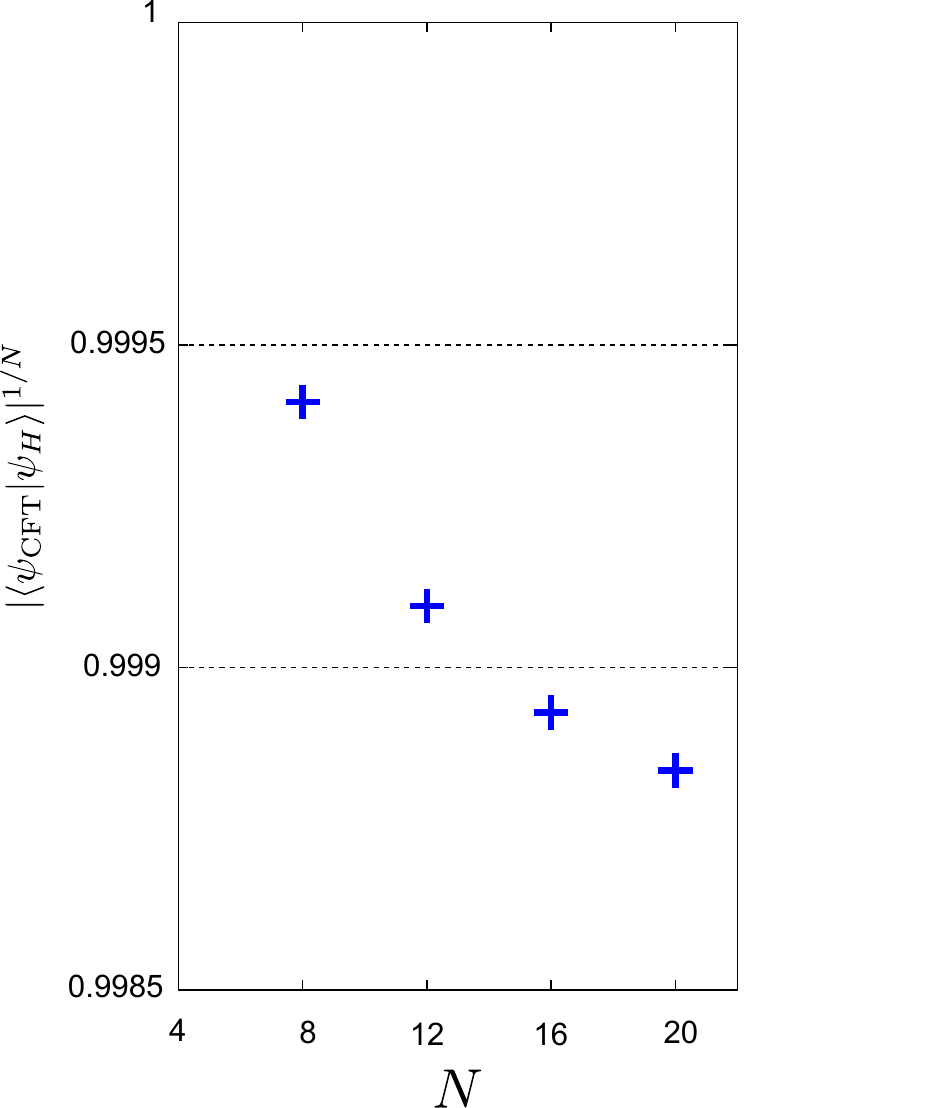}\label{fig:overlap1D2}}
\caption{(a) Overlap $|\Braket{\psi_{\text{CFT}}|\psi_H }|$ between the $(q=2,\eta=1)_{1/2}$ CFT state and the ground state of Hamiltonian $H^{(2)}_{1D}$ with $\kappa$ fixed to $0.274+0.052 i$, as a function of the number $N$ of lattice sites.  The dotted line is a linear fit with equation $y=1.0075-0.00152 N$. If the overlap continues to follow this behavior at larger sizes, it will still be above $85\%$ for a spin chain with 100 lattice sites. (b) Overlap per site $|\Braket{\psi_{\text{CFT}}|\psi_H }|^{1/N}$ between the same two states.}
\end{center}
\end{figure}

In two dimensions, since the $SU(2)$ symmetry is not present in this model, cutting the parent Hamiltonian leads to a local Hamiltonian with up to five-body interactions with many different coefficients. In addition, the fact that the five-body terms involve more sites means that each of them stretches over a larger part of the lattice. With the limited lattice sizes that we can consider with exact diagonalization, this is problematic because the local regions need to be small compared to the total size of the lattice (otherwise it would not be expected that the same local Hamiltonian would also work for other lattice sizes). This suggests that even if a local Hamiltonian that is related to the exact Hamiltonian exists, we may not be able to find it with exact diagonalization. Instead
of cutting the exact parent Hamiltonian, we therefore asked whether, by chance, a local Hamiltonian can be obtained if we restrict the range of the interactions to all interactions preserving the number of particles on all possible configurations inside a plaquette of the lattice. Optimizing the coefficients in this Hamiltonian, we did not, however, find a set of coefficients for these interactions for which the ground state of this Hamiltonian is close to the $(q = 2, \eta = 1)_{1/2}$ CFT state. Whether there exists a more complicated, but still local, Hamiltonian stabilizing this state therefore remains an interesting open problem.

\section{Conclusion}
\label{SEC:Conclusion}
We have introduced a three-parameter family $(q,\eta)_S$ of strongly-correlated spin states on arbitrary lattices in one and two dimensions. It was shown that these states reduce to continuum Moore-Read states of bosons (odd q) and fermions (even q) in the continuum limit $\eta\rightarrow 0$. Numerical evidence that these states are critical states in one dimension and topological states in two dimensions was provided, and the topological entanglement entropy was shown to remain the same along the interpolation for the $(q=1,\eta)_1$ and $(q=2,\eta)_{1/2}$ states. Parent Hamiltonians of the states in the lattice limit were derived using analytical tools from CFT and in some cases it was shown that these states could be stabilized by local Hamiltonians in one and two dimensions.

There is currently a lot of interest in finding models possessing topological properties. Given the complexity of quantum many-body systems, the analysis of phenomena, like e.g.\ topology, can be greatly facilitated by having models in which at least the ground state can be found analytically. The results of the present paper show that CFT is a valuable tool to derive analytical models, also in the context of non-Abelian FQH states, and in addition can be used as a starting point to identify simpler models that are more realistic to realize physically.

\begin{acknowledgments}
The authors would like to thank Hong-Hao Tu for discussions on related topics. This work was supported by the EU integrated project SIQS, FIS2012-33642, QUITEMAD+ (CAM), the Severo Ochoa Program and the Fulbright grant PRX14/00352.
\end{acknowledgments}

\allowdisplaybreaks
\appendix
\section{Operators annihilating the $(q,\eta=1)_1$ CFT wave functions}
\label{appendixA}

We restrict ourselves to the case $q\geq 2$, since the construction of the parent Hamiltonian when $q=1$ is explained in \ref{parentq1}. Note however that the derivation presented here can also be used to obtain a Hamiltonian without the $SU(2)$ symmetry for the $(q=1,\eta=1)_1$ CFT state. The CFT states are defined from the operators $\mathcal{V}_{n_j}(z_j)$ given in Eq. (\ref{defvertex}). Additional operators that are needed to construct parent Hamiltonians are the operators $G^{\pm }(z)=\chi(z) e^{\pm i\sqrt{q}\phi (z)}$, $J(z)=\frac{i}{\sqrt{q}}\partial \phi (z)$. The operator product expansion (OPE) of $G^{+}(z)$ and $J(z)$ with the operators used to build the wave function are, for $q\geq 2$,
\begin{align}
G^{+ }(z)V_{n_{j}}(w)&\sim (-1)^{j-1} \frac{\delta_{n_j,0}}{z-w} \mathcal{V}_{1}(w),\\
J(z)V_{n_{j}}(w)&\sim \frac{1}{q}\frac{qn_j-1}{z-w} \mathcal{V}_{n_{j}}(w).
\end{align}
We first show that $\chi^{q}(w)$ is a null field, where
\begin{align}
\chi^{q}(w) &=\oint_{w}\frac{dz}{2\pi i}\frac{1}{z-w} \frac{1}{(z-w)^{q-1}}G^{+}(z)V_{1}(w) \nonumber \\
&\ \ \ - \oint_{w}\frac{dz}{2\pi i}\frac{1}{z-w} qJ(z)V_{2}(w),\\
&=\Omega^q_{2}(w) - \Omega^q_{3}(w),
\end{align}
where the contour of the integration is a circle around $w$ traversed counter-clockwise. We have that
\begin{align}
\Omega^q_{2}(w) &=\oint_{w}\frac{dz}{2\pi i}\frac{1}{z-w}\left[\frac{1}{(z-w)^{q-1}}G^{+}(z)V_{1}(w)\right],\nonumber\\
&=\oint_{w}\frac{dz}{2\pi i}\frac{1}{z-w}\left[\frac{1}{(z-w)^{q-1}}\chi(z)\chi(w)\right.\nonumber\\
& \left. \qquad \qquad \qquad \times e^{+ i\sqrt{q}\phi (z)}e^{i(q-1)\phi (w)/\sqrt{q}}\right],\nonumber\\
&=\oint_{w}\frac{dz}{2\pi i}\frac{1}{z-w}\left[\frac{(z-w)^{q-1}}{(z-w)^{q-1}}\chi(z)\chi(w)\right.\nonumber\\
& \left. \qquad \qquad \qquad \times e^{i\sqrt{q}\phi (z)+i(q-1)\phi (w)/\sqrt{q}}\right],\nonumber\\
&=\oint_{w}\frac{dz}{2\pi i}\frac{1}{z-w}\left[\left(\frac{1}{z-w}+(z-w)A(w)+...\right)\right.\nonumber\\
& \left. \qquad \qquad \qquad \times e^{i\sqrt{q}\phi (z)+i(q-1)\phi (w)/\sqrt{q}}\right],\nonumber\\
&=\oint_{w}\frac{dz}{2\pi i}\left[\left(\frac{1}{(z-w)^2}+A(w)+...\right)\right.\nonumber\\
& \left. \qquad \qquad \qquad \times e^{i\sqrt{q}\phi (z)+i(q-1)\phi (w)/\sqrt{q}}\right],\nonumber\\
&=\oint_{w}\frac{dz}{2\pi i}\left[\frac{1}{(z-w)^2}e^{i\sqrt{q}\phi (z)+i(q-1)\phi (w)/\sqrt{q}}\right],\nonumber\\
&=\oint_{w}\frac{dz}{2\pi i} \frac{1}{z-w} \left[i \sqrt{q} \partial \phi (w) e^{i(2q-1)\phi (w)/\sqrt{q}}\right],
\end{align}
and
\begin{align}
\Omega^q_{3}(w) &=\oint_{w}\frac{dz}{2\pi i}\frac{1}{z-w}\left[qJ(z)V_{2}(w)\right]\\
&=\oint_{w}\frac{dz}{2\pi i}\frac{1}{z-w}\left[\sqrt{q} i \partial \phi (z) e^{i(2q-1)\phi (w)/\sqrt{q}}\right]\\
&=\oint_{w}\frac{dz}{2\pi i}\frac{1}{z-w}\left[\sqrt{q} i \partial \phi (w) e^{i(2q-1)\phi (w)/\sqrt{q}}\right]\\
&=\Omega^q_{2}(w),
\end{align}
which shows that $\chi^{q}(w)$ is a null field. Similarly, there are more simple null fields 
 \begin{align}
 \underset{p=0,1,\ldots,q-2}{\chi^{p}(w)} &=\oint_{w}\frac{dz}{2\pi i} \frac{1}{(z-w)^{p}} G^{+}(z) V_{1}(w),\\
 \chi^{q-1}(w) &=\oint_{w}\frac{dz}{2\pi i}\left[\frac{1}{(z-w)^{q-1}} G^{+}(z) V_{1}(w)\right.\nonumber\\
 &\left. \ \ \ \ \ \ \ \ \ \ -\frac{1}{(z-w)}V_{2}(w)\right].
 \end{align}
 
 Let us now use the fact that by replacing the field at site $i$ by a null field, the correlator vanishes :
 \begin{align}
 0 &=\langle \mathcal{V}_{n_{1}}(z_{1})\cdots \chi^a(z_{i})\cdots
 \mathcal{V}_{n_{N}}(z_{N})\rangle.
 \end{align}
 We will transform this equation into an equation involving the wave function by deforming the contour integral and moving the operators in the null fields at different positions. Let us do it for the null field $\chi^{q}(w)$. We will use the OPEs as well as the commutation relations :
 \begin{equation}
 \mathcal{V}_{n_j}(z_j)G^{+}(z)=(-1)^{(q+1)n_j-1}G^{+}(z)\mathcal{V}_{n_j}(z_j),
 \end{equation}
 where we have used $:e^{i\alpha\phi(z)}::e^{i\beta\phi(w)}:{}=
 (-1)^{\alpha\beta}:e^{i\beta\phi(w)}::e^{i\alpha\phi(z)}:$ and $\chi(z)\chi(w)=(-1)\chi(w)\chi(z)$, which adds a minus sign only when $n_j=1$.\\
 
 We then have, starting with the first term involving $\chi^q_{2}(w)$ :
 
  \begin{align}
  &\oint_{z_i}\frac{dz}{2\pi i}\frac{1}{(z-z_i)^q}
  \langle \mathcal{V}_{n_{1}}(z_{1})\ldots G^{+}(z)V_{1}(z_i) \ldots \mathcal{V}_{n_{N}}(z_{N})\rangle,\nonumber\\
  &=-\sum_{j(\neq i)}\oint_{z_j}\frac{dz}{2\pi i}\frac{1}{(z-z_i)^q}\nonumber\\
 &\qquad \times  \ \langle \mathcal{V}_{n_{1}}(z_{1})\ldots G^{+}(z)V_{1}(z_i)\ldots \mathcal{V}_{n_{N}}(z_{N})\rangle,\nonumber\\
  &=-(-1)^{i-1}\sum_{j=1}^{i-1}\oint_{z_j}\frac{dz}{2\pi i}\frac{(-1)^{(q+1)\sum_{k=j}^{i-1}n_k}}{(z-z_i)^q}
 \frac{\delta_{n_j,0}}{z-z_j}\nonumber\\
 &\qquad \times  \langle \mathcal{V}_{n_{1}}(z_{1})\ldots \mathcal{V}_{1}(z_j) \ldots V_{1}(z_i) \ldots \mathcal{V}_{n_{N}}(z_{N})\rangle\nonumber\\
  &\phantom{=}-(-1)^{i-1}\sum_{j=i+1}^N\oint_{z_j}\frac{dz}{2\pi i}\frac{(-1)^{(q+1)}(-1)^{(q+1)\sum_{k=i+1}^{j-1}n_k}}{(z-z_i)^q} \nonumber\\
  &\qquad \times \frac{\delta_{n_j,0}}{z-z_j} \langle \mathcal{V}_{n_{1}}(z_{1})\ldots V_{1}(z_i) \ldots \mathcal{V}_{1}(z_j) \ldots \mathcal{V}_{n_{N}}(z_{N})\rangle,\nonumber\\
  &=-(-1)^{i-1}\sum_{j=1}^{i-1}\frac{(-1)^{(q+1)\sum_{k=j}^{i-1}n_k}}{(z_j-z_i)^q}
  \delta_{n_j,0} \nonumber\\
  &\qquad \times  \langle \mathcal{V}_{n_{1}}(z_{1})\ldots \mathcal{V}_{1}(z_j) \ldots V_{1}(z_i) \ldots \mathcal{V}_{n_{N}}(z_{N})\rangle\nonumber\\
  &\phantom{=}-(-1)^{i-1}\sum_{j=i+1}^N\frac{(-1)^{(q+1)}(-1)^{(q+1)\sum_{k=i+1}^{j-1}n_k}}{(z_j-z_i)^q}
  \delta_{n_j,0} \nonumber\\
  &\qquad \times  \langle 
  \mathcal{V}_{n_{1}}(z_{1})\ldots V_{1}(z_i) \ldots \mathcal{V}_{1}(z_j) \ldots \mathcal{V}_{n_{N}}(z_{N})\rangle\nonumber,\\
  &=-\sum_{j=1}^{i-1}\frac{(-1)^{(q+1)\sum_{k=j+1}^{i-1}n_k}}{(z_j-z_i)^q} \delta_{n_j,0}\nonumber\\
  &\qquad \times  \psi_{(q,\eta=1)_1}(n_1,\ldots,1,\ldots,1,\ldots,n_N)\nonumber\\
  &\phantom{=}-\sum_{j=i+1}^N\frac{(-1)^{(q+1)}(-1)^{(q+1)\sum_{k=i+1}^{j-1}n_k}}{(z_j-z_i)^q} \delta_{n_j,0}\nonumber\\
  &\qquad \times  \psi_{(q,\eta=1)_1}(n_1,\ldots,1,\ldots,1,\ldots,n_N),\nonumber \\
    &=\sum_{j=1}^{i-1}(-1)^{(q+1)}\frac{(-1)^{(q+1)\sum_{k=j+1}^{i-1}n_k}}{(z_i-z_j)^q}\delta_{n_j,0} \nonumber\\
      &\qquad \times  \psi_{(q,\eta=1)_1}(n_1,\ldots,1,\ldots,1,\ldots,n_N)\nonumber\\
    &\phantom{=}+\sum_{j=i+1}^N\frac{(-1)^{(q+1)\sum_{k=i+1}^{j-1}n_k}}{(z_i-z_j)^q} \delta_{n_j,0}\nonumber\\
    &\qquad \times  \psi_{(q,\eta=1)_1}(n_1,\ldots,1,\ldots,1,\ldots,n_N),\label{equationnullfield}
  \end{align}
 where $\psi_{(q,\eta=1)_1}$ is the wave function of the $(q,\eta=1)_1$ CFT state.
 
 Let us now define the creation and annihilation operators $d_j, d_j^\dagger, d_i', d_i'^\dagger$ acting on the Hilbert space at site j as
 \begin{align}
d_j \Ket{n_j}&=(-1)^{(q+1)\sum_{k=1}^{j-1} n_k} \begin{cases}0 & n_j=0 \\
\Ket{0} & n_j=1\\
0 & n_j=2
\end{cases}\\
d_j^\dagger \Ket{n_j}&=(-1)^{(q+1)\sum_{k=1}^{j-1} n_k}\begin{cases}\Ket{1} & n_j=0 \\
0 & n_j=1\\
0 & n_j=2
\end{cases}\\
d_j' \Ket{n_j}&=(-1)^{(q+1)\sum_{k=1}^{j-1} n_k}\begin{cases}0 & n_j=0 \\
0 & n_j=1\\
\Ket{1} & n_j=2
\end{cases}\\
d_j'^\dagger \Ket{n_j}&=(-1)^{(q+1)\sum_{k=1}^{j-1} n_k}\begin{cases}0 & n_j=0 \\
\Ket{2} & n_j=1\\
0 & n_j=2
\end{cases}
 \end{align}
 We also define the particle number operators corresponding to these operators as $n_j^{(1)}=d_j^\dagger d_j$ and $n_j^{(2)}=d_j'^\dagger d_j'$, such that $n_j=n_j^{(1)}+2n_j^{(2)}$.
 
 We multiply Eq.~(\ref{equationnullfield}) by 
 \begin{align}
 &|n_1,\ldots,n_{j-1},n_j,n_{j+1}\ldots,n_{i-1},2,n_{i+1}\ldots,n_N\rangle,
 \end{align}
 and sum over all $n_k$, $k\neq i$, to get an expression involving the wave function
 \begin{equation}
 \sum_{j(\neq i)}\frac{1}{(z_i-z_j)^q} d_j d_i'^\dagger|\psi_{(q,\eta=1)_1}\rangle.
 \end{equation}
 
 Let us now look at the second term involving $\chi^q_{3}(w)$:
 \begin{align}
 &-q\oint_{z_i}\frac{dz}{2\pi i}\frac{1}{z-z_i}\langle
 \mathcal{V}_{n_{1}}(z_{1})\ldots J(z)V_2(z_i) \ldots \mathcal{V}_{n_{N}}(z_{N})\rangle,\\
 &=q\sum_{j(\neq i)}\oint_{z_j}\frac{dz}{2\pi i}\frac{1}{z-z_i}\langle
 \mathcal{V}_{n_{1}}(z_{1})\ldots J(z)V_2(z_i) \ldots \mathcal{V}_{n_{N}}(z_{N})\rangle,\\
 &=\sum_{j(\neq i)}\frac{qn_j-1}{z_j-z_i}\langle
 \mathcal{V}_{n_{1}}(z_{1}) \ldots V_2(z_i) \ldots \mathcal{V}_{n_{N}}(z_{N})\rangle.
 \end{align}
 Let us multiply this expression by $|n_1,\ldots,n_{i-1},2,n_{i+1}\ldots,n_N\rangle =\sum_{n'_i}n_i^{(2)}|n_1,\ldots,n'_i,\ldots,n_N\rangle$,
 which leads after summing over all $n_k$, $k\neq i$ to
 \begin{equation}\label{secondterm}
 -\sum_{j(\neq i)}\frac{qn_j-1}{z_i-z_j}n^{(2)}_i|\Psi\rangle.
 \end{equation}

 Summing the two terms together, we get an operator annihilating the wave function :
 \begin{equation}
 \Lambda_i^{q}=\sum_{j(\neq i)}\frac{1}{(z_i-z_j)^q}d_j d_i'^\dagger-\sum_{j(\neq i)}\frac{qn_j-1}{z_i-z_j}n_i^{(2)}.
 \end{equation}
 
 The same procedure applied to the other null fields gives the following operators annihilating the $(q,\eta=1)_1$ CFT wave functions :
 \begin{align}
 \Lambda^{0} &=\sum_i d_i,\\
 \underset{p=1,\ldots,q-2}{\Lambda_i^{p}} &=\sum_{j(\neq i)} \frac{1}{(z_i-z_j)^{p}} d_j d_i'^\dagger,\\
 \Lambda_i^{q-1} &=\sum_{j(\neq i)} \frac{1}{(z_i-z_j)^{q-1}} d_j d_i'^\dagger+n_i^{(2)}.
 \end{align}
 
 \section{Operators annihilating the $(q,\eta=1)_{1/2}$ CFT wave functions}
 \label{appendixB}
 
 To obtain operators annihilating the $(q,\eta=1)_{1/2}$ CFT wave functions, we can follow the same procedure and use the following null field instead of $\chi^{q}(w)$ :
 \begin{align}
&\oint_{w}\frac{dz}{2\pi i}\frac{1}{(z-w)^q} G^{+}(z) \left( \oint_{w}\frac{dx}{2\pi i} G^{+}(x)V_{0}(w)\right)\nonumber\\
&\ \ -q\oint_{w}\frac{dz}{2\pi i}\frac{1}{z-w} J(z) \left( \oint_{w}\frac{dx}{2\pi i}\frac{1}{(x-w)^{q-1}} G^{+}(x)V_{1}(w)\right).
 \end{align}
 
However this procedure does not work when $q=2$ since this is not a null field when $q=2$. Instead of following this approach, we present a different way to obtain operators annihilating the $(q,\eta=1)_{1/2}$ CFT wave functions, using the operators already obtained in Appendix \ref{appendixA}. The resulting operators are the same that would be obtained directly using null fields when $q>2$ but this approach allows us to also construct a parent Hamiltonian when $q=2$.
 
 We start by multiplying the previously obtained operators on the left by $d_i'$. Since $d_i' d_i'^\dagger = n_i^{(1)}$, this leads to new operators
  \begin{align}
  \underset{p=1,\ldots,q-2}{\Lambda_i^{''p}} &=\sum_{j(\neq i)} \frac{1}{(z_i-z_j)^{p}} d_j n_i^{(1)},\\
  \Lambda_i^{''q-1} &=\sum_{j(\neq i)} \frac{1}{(z_i-z_j)^{q-1}} d_j n_i^{(1)}+d_i',\\
  \Lambda_i^{''q}&=\sum_{j(\neq i)}\frac{1}{(z_i-z_j)^q}d_j n_i^{(1)}-\sum_{j(\neq i)}\frac{qn_j-1}{z_i-z_j}d_i'.
  \end{align}
 The operator $d_i'$ in $\Lambda_i^{''q}$ can be replaced by $-\sum_{h(\neq i)} \frac{1}{(z_i-z_h)^{q-1}} d_h n_i^{(1)}$ since $\Lambda_i^{''q-1}$ annihilates the $(q,\eta=1)_{1}$ wave function :
 \begin{align}
   \Lambda_i^{''q}=&\sum_{j(\neq i)}\frac{1}{(z_i-z_j)^q}d_j n_i^{(1)}\nonumber\\
   &+\sum_{j(\neq i)}\sum_{h(\neq i)} \frac{1}{(z_i-z_h)^{q-1}}\frac{qn_j^{(1)}-1}{z_i-z_j} d_h n_i^{(1)}.
   \end{align}
 
 $\Lambda_i^{''q}$ then acts separately on $\mathcal{H}^{1}$ and on $\mathcal{H}^{2}$ and the operators $\Lambda_i^{''1},\ldots,\Lambda_i^{''q-2}$ give zero on $\mathcal{H}^{2}$. By keeping only the terms acting on $\mathcal{H}^{1}$ and since the $(q,\eta=1)_{1/2}$ states are projections of the $(q,\eta=1)_{1}$ states on $\mathcal{H}^{1}$, we get operators annihilating the $(q,\eta=1)_{1/2}$ wave functions :
 
 \begin{align}
 \Lambda^{'0} &=\sum_i d_i,\\
 \underset{p=1,\ldots,q-2}{\Lambda_i^{'p}} &=\sum_{j(\neq i)} \frac{1}{(z_i-z_j)^{p}} d_j n_i^{(1)},\\
 \Lambda_i^{'q-1} &=\sum_{j(\neq i)}\frac{1}{(z_i-z_j)^q}d_j n_i^{(1)}\nonumber\\
    &\ \ \ +\sum_{j(\neq i)}\sum_{h(\neq i)} \frac{1}{(z_i-z_h)^{q-1}}\frac{qn_j^{(1)}-1}{z_i-z_j} d_h n_i^{(1)}.
 \end{align}

\bibliography{biblio}
\end{document}

%% file: interpolationtest.pdf_tex
%% Creator: Inkscape 0.48.3.1, www.inkscape.org
%% Accompanies image file 'interpolationtest.pdf' (pdf, eps, ps)
\begingroup%
  \makeatletter%
  \providecommand\color[2][]{%
    \errmessage{(Inkscape) Color is used for the text in Inkscape, but the package 'color.sty' is not loaded}%
    \renewcommand\color[2][]{}%
  }%
  \providecommand\transparent[1]{%
    \errmessage{(Inkscape) Transparency is used (non-zero) for the text in Inkscape, but the package 'transparent.sty' is not loaded}%
    \renewcommand\transparent[1]{}%
  }%
  \providecommand\rotatebox[2]{#2}%
  \ifx\svgwidth\undefined%
    \setlength{\unitlength}{855.98066406bp}%
    \ifx\svgscale\undefined%
      \relax%
    \else%
      \setlength{\unitlength}{\unitlength * \real{\svgscale}}%
    \fi%
  \else%
    \setlength{\unitlength}{\svgwidth}%
  \fi%
  \global\let\svgwidth\undefined%
  \global\let\svgscale\undefined%
  \makeatother%
  \begin{picture}(1,0.38567619)%
    \put(0,0){\includegraphics[width=\unitlength]{interpolationtest.pdf}}%
    \put(0.32527587,0.16905951){\color[rgb]{0,0,0}\makebox(0,0)[lb]{\smash{$\text{Re}(z_j)$}}}%
    \put(0.14609404,0.37038579){\color[rgb]{0,0,0}\makebox(0,0)[lb]{\smash{$\text{Im}(z_j)$}}}%
    \put(0.68816235,0.37038578){\color[rgb]{0,0,0}\makebox(0,0)[lb]{\smash{$\text{Im}(z_j)$}}}%
    \put(0.86921342,0.16905951){\color[rgb]{0,0,0}\makebox(0,0)[lb]{\smash{$\text{Re}(z_j)$}}}%
    \put(0.78404641,0.00332221){\color[rgb]{0,0,0}\makebox(0,0)[lb]{\smash{$a=2\pi$}}}%
    \put(0.19846614,0.01065886){\color[rgb]{0,0,0}\makebox(0,0)[lb]{\smash{$a\rightarrow 0^+$}}}%
    \put(0.35558562,0.22497164){\color[rgb]{0,0,0}\makebox(0,0)[lb]{\smash{$0\leftarrow\eta\rightarrow1$}}}%
  \end{picture}%
\endgroup%

%% file: cylinder2bien.pdf_tex
%% Creator: Inkscape 0.48.3.1, www.inkscape.org
%% Accompanies image file 'cylinder2bien.pdf' (pdf, eps, ps)
\begingroup%
  \makeatletter%
  \providecommand\color[2][]{%
    \errmessage{(Inkscape) Color is used for the text in Inkscape, but the package 'color.sty' is not loaded}%
    \renewcommand\color[2][]{}%
  }%
  \providecommand\transparent[1]{%
    \errmessage{(Inkscape) Transparency is used (non-zero) for the text in Inkscape, but the package 'transparent.sty' is not loaded}%
    \renewcommand\transparent[1]{}%
  }%
  \providecommand\rotatebox[2]{#2}%
  \ifx\svgwidth\undefined%
    \setlength{\unitlength}{387.50264796bp}%
    \ifx\svgscale\undefined%
      \relax%
    \else%
      \setlength{\unitlength}{\unitlength * \real{\svgscale}}%
    \fi%
  \else%
    \setlength{\unitlength}{\svgwidth}%
  \fi%
  \global\let\svgwidth\undefined%
  \global\let\svgscale\undefined%
  \makeatother%
  \begin{picture}(1,1.24113666)%
    \put(0,0){\includegraphics[width=\unitlength]{cylinder2bien.pdf}}%
    \put(0.70134075,0.84961787){\color[rgb]{0,0,0}\makebox(0,0)[lb]{\smash{$\text{Re}(z_j)$}}}%
    \put(0.36325196,1.20736069){\color[rgb]{0,0,0}\makebox(0,0)[lb]{\smash{$\text{Im}(z_j)$}}}%
    \put(0.89401484,0.30774638){\color[rgb]{0,0,0}\makebox(0,0)[lb]{\smash{$L_y$}}}%
    \put(0.33858075,0.0080947){\color[rgb]{0,0,0}\makebox(0,0)[lb]{\smash{$L_x$}}}%
  \end{picture}%
\endgroup%

%% file: gapplane2.pdf_tex
%% Creator: Inkscape 0.48.3.1, www.inkscape.org
%% Accompanies image file 'gapplane2.pdf' (pdf, eps, ps)
\begingroup%
  \makeatletter%
  \providecommand\color[2][]{%
    \errmessage{(Inkscape) Color is used for the text in Inkscape, but the package 'color.sty' is not loaded}%
    \renewcommand\color[2][]{}%
  }%
  \providecommand\transparent[1]{%
    \errmessage{(Inkscape) Transparency is used (non-zero) for the text in Inkscape, but the package 'transparent.sty' is not loaded}%
    \renewcommand\transparent[1]{}%
  }%
  \providecommand\rotatebox[2]{#2}%
  \ifx\svgwidth\undefined%
    \setlength{\unitlength}{505.16113281bp}%
    \ifx\svgscale\undefined%
      \relax%
    \else%
      \setlength{\unitlength}{\unitlength * \real{\svgscale}}%
    \fi%
  \else%
    \setlength{\unitlength}{\svgwidth}%
  \fi%
  \global\let\svgwidth\undefined%
  \global\let\svgscale\undefined%
  \makeatother%
  \begin{picture}(1,0.70254608)%
    \put(0,0){\includegraphics[width=\unitlength]{gapplane2.pdf}}%
    \put(0.51041006,0.01349747){\color[rgb]{0.14901961,0.14901961,0.14901961}\makebox(0,0)[b]{\smash{$1/N$}}}%
    \put(0.03185712,0.35358246){\color[rgb]{0.14901961,0.14901961,0.14901961}\rotatebox{90}{\makebox(0,0)[b]{\smash{$E_n-E_0$}}}}%
    \put(0.11049015,0.03857303){\color[rgb]{0,0,0}\makebox(0,0)[b]{\smash{\tiny 0}}}%
    \put(0.37809268,0.03857303){\color[rgb]{0,0,0}\makebox(0,0)[b]{\smash{\tiny 0.05}}}%
    \put(0.64569521,0.03857303){\color[rgb]{0,0,0}\makebox(0,0)[b]{\smash{\tiny 0.1}}}%
    \put(0.09443349,0.11342598){\color[rgb]{0,0,0}\makebox(0,0)[rb]{\smash{\tiny 0}}}%
    \put(0.09443349,0.21836675){\color[rgb]{0,0,0}\makebox(0,0)[rb]{\smash{\tiny 2}}}%
    \put(0.09443349,0.32330752){\color[rgb]{0,0,0}\makebox(0,0)[rb]{\smash{\tiny 4}}}%
    \put(0.09443349,0.42824829){\color[rgb]{0,0,0}\makebox(0,0)[rb]{\smash{\tiny 6}}}%
    \put(0.09443349,0.53318906){\color[rgb]{0,0,0}\makebox(0,0)[rb]{\smash{\tiny 8}}}%
    \put(0.09443349,0.63812983){\color[rgb]{0,0,0}\makebox(0,0)[rb]{\smash{\tiny 10}}}%
    \put(0.24320783,0.61983748){\color[rgb]{0,0,0}\makebox(0,0)[b]{\smash{{\color{red} \tiny $L_x\times L_y$ :}}}}%
    \put(0.43958081,0.61983748){\color[rgb]{0,0,0}\makebox(0,0)[b]{\smash{{\color{red} \tiny $4\times 4$}}}}%
    \put(0.55360383,0.61983748){\color[rgb]{0,0,0}\makebox(0,0)[b]{\smash{{\color{red} \tiny $4\times 3$}}}}%
    \put(0.77848262,0.61983748){\color[rgb]{0,0,0}\makebox(0,0)[b]{\smash{{\color{red} \tiny $4\times 2$}}}}%
    \put(0.23788499,0.08866814){\color[rgb]{0,0,0}\makebox(0,0)[b]{\smash{{\color{OliveGreen} \tiny Overlap :}}}}%
    \put(0.44097231,0.08997182){\color[rgb]{0,0,0}\makebox(0,0)[b]{\smash{{\color{OliveGreen} \tiny $97.36 \%$}}}}%
    \put(0.55499533,0.08997182){\color[rgb]{0,0,0}\makebox(0,0)[b]{\smash{{\color{OliveGreen} \tiny $98.10 \%$}}}}%
    \put(0.77987411,0.08997182){\color[rgb]{0,0,0}\makebox(0,0)[b]{\smash{{\color{OliveGreen} \tiny $99.18 \%$}}}}%
  \end{picture}%
\endgroup%